\documentclass[12pt,a4paper,titlepage]{article}

\usepackage{graphicx}
\usepackage{dcolumn}
\usepackage{bm}
\usepackage{nicefrac}
\usepackage{float}
\usepackage{eqnarray,amsmath}
\usepackage{subcaption}
\pdfoutput=1

\begin{document}

\title{Drop impact on wetted walls: An analytical solution for modelling the crown spreading based on stagnation-point flow}

\author{G. Lamanna, A. Geppert, R. Bernard, I. H\"{o}rner, B. Weigand\\
Institute of Aerospace Thermodynamics,\\
University of Stuttgart,\\
Pfaffenwaldring 31,\\ 70569 Stuttgart, Germany.\\
Email: Grazia.Lamanna@itlr.uni-stuttgart.de,\\ Anne.Geppert@itlr.uni-stuttgart.de}

\maketitle

\begin{abstract}
An analytical solution is proposed to predict the crown propagation, generated by a single droplet impact on wetted walls. This approach 
enables a smooth transition from the inertia-driven to the viscous-controlled regime of crown propagation. The modelling strategy is based 
on the stagnation-point flow, because it resembles closely the hydrodynamic flow in the lamella and offers two main advantages. First, it 
allows a simple estimation of the wall-film thinning rate, caused by the impulse transfer from the impacting droplet to the wall film. Second, 
thanks to the self-similarity of the solution, it enables a straightforward estimation of momentum losses during film spreading along the wall. 
By incorporating this estimation into existing inviscid models, an excellent agreement with experiments is found during the entire crown 
elevation phase. In general, the analysis shows that momentum losses due to viscous effects cannot be neglected during a significant portion 
of crown propagation, particularly for thin wall films. The proposed methodology paves the way for predicting the inception of crown bottom 
breakup (CBB). In this case, the crown lamella disintegrates directly at its base due to the spontaneous creation of holes that create a web-like structure in the lamella prior to its break-up. Our theoretical analysis shows that this premature break-up of the crown lamella is associated 
to local instability effects, caused by the unbalance between inertial forces and surface tension.
\end{abstract}


\section{\label{sec:Intro}Introduction}
\begin{figure}
\centering
{\includegraphics[width=0.5\columnwidth]{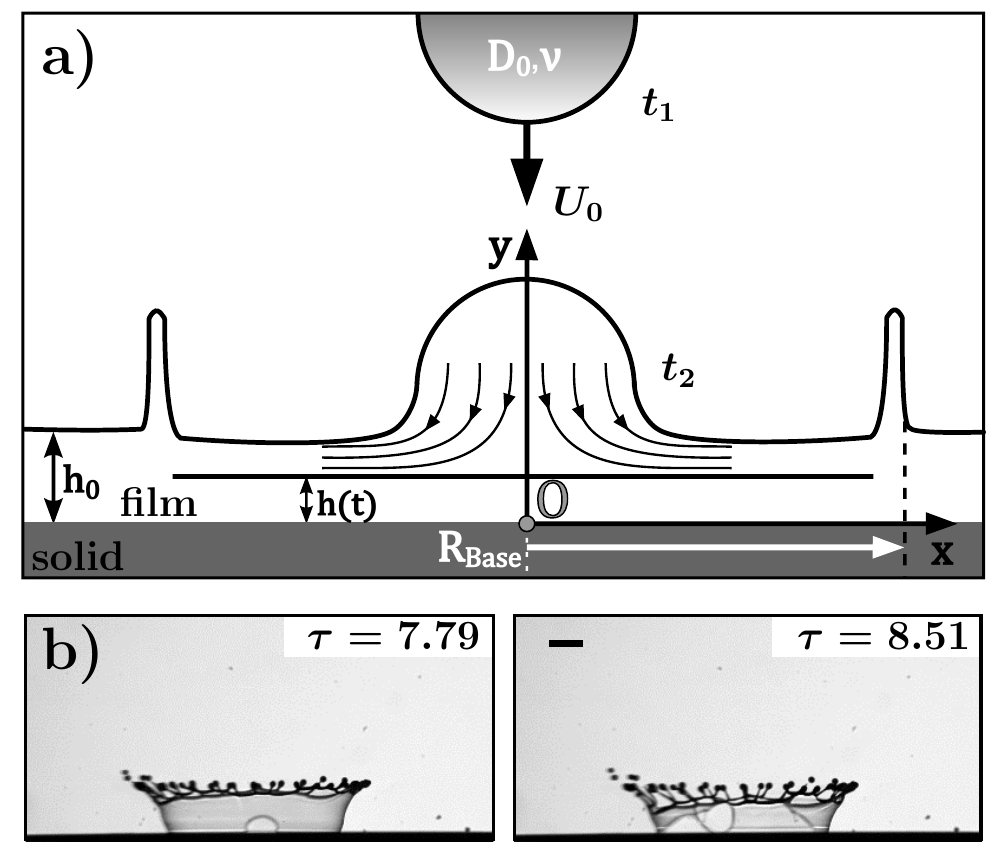}} 
\caption{(a) Sketch of the droplet impact flow against a wetted wall and (b) crown bottom breakup time sequence. Fluids: hyspin for droplet 
and wall film. $We_0 = 1400$, $\delta = 0.05$. The exact specifications of test fluid and impact conditions can be found in Table \ref{tab:Exp} 
(see Appendix \ref{app:TC}).}
\label{fig:Intro1}
\end{figure}

Drop impact on wetted surfaces is of pertinence to many technical applications as well as to natural sciences, such as soil erosion, 
IC engines, icing on plane wings and spray coating technologies. 
Immediately after the impact, the droplet expands radially along the surface. If the impact kinetic energy is sufficiently high to 
overcome energy losses due to deformation and viscous effects, an upward growing crown is generated with detachment of secondary 
droplets (splashing regime). It is widely accepted that the non-dimensional parameter $K$ represents a meaningful choice for predicting 
the onset of the splashing regime. The latter can be expressed as a function of Reynolds ($Re_0 = U_0 D_0 / \nu$) and Weber 
($We_0 = \rho U^2_0 D_0 / \sigma$) numbers as $K = We_0^{0.5} Re_0^{0.25}$. Here $\nu$, $\rho$, and $\sigma$ are the kinematic 
viscosity, density, and surface tension of the impacting liquid droplet and $U_0$ and $D_0$ its velocity and diameter, respectively. In 
literature, several empirical correlations can be found that express the threshold parameter $K$ in terms of the non-dimensional film 
thickness $\delta = h_0 /D_0$, both for one-component \cite{VanderWal2006b,Okawa2006,Yarin1995,Gao2015,Tropea1999,Cossali1997} 
and two-components \cite{Geppert2016,Kittel2018} interactions. Here $h_0$ denotes the initial film thickness, as indicated in 
Fig.~\ref{fig:Intro1}a.
The determination of the splashing threshold alone does not provide an exhaustive picture of the complex splashing dynamics. An example 
is shown in Fig.~\ref{fig:Intro1}b, illustrating the splashing sequence of a hyspin (hydraulic oil) droplet impinging on a very thin wall film of 
the same liquid ($\delta = 0.05$). As can be seen, the crown is ripped off at its base already during the early stage of the elevation phase 
($\tau = t \, U_0 /D_0 = 7.79$). This phenomenon of crown bottom breakup (CBB) is generally observed for very thin films, provided 
$\delta < 0.08$. Indeed, it has been observed in \cite{Geppert2016a} for \textit{n}-hexadecane/hyspin interactions, in \cite{Abouelsoud2018} 
for ethylene glycol, in \cite{Wang2000} for glycerol-water solutions and in \cite{Tropea1999} for isopropanol. Here, we postulate that CBB is  
connected to a significant increase of momentum losses during the spreading of the lamella along a thinning wall film. Indeed, if the mass	 
flow rate approaches zero during the stretching phase of the crown, the lamella is no longer capable to sustain the crown growth. If this 
occurs, the crown breaks up at its foot. 

The importance of viscous losses on the crown propagation has been recently highlighted by Marcotte et al.~\cite{Marcotte2019}. Their 
simulations show that the crown spreading consists out of two distinct sheets, originating from the liquid drop and the substrate liquid, 
respectively. The two sheets evolve on separate time-scales, and their merging time is mainly depending upon the viscosity ratio. These 
findings imply that the rate of crown propagation cannot be correctly predicted without including an accurate estimation of viscous losses 
during the spreading of the lamella, as currently assumed by the majority of theoretical models, see e.g.~\cite{Yarin1995,Cossali2004,Davidson2002,Gao2015}. In all cases, the starting point is represented by the pioneering work of 
Yarin and Weiss~\cite{Yarin1995}, who modelled the crown as a kinematic discontinuity that propagates with the square root of time 
$R_{Base}/D_0 = C_1 \sqrt{\tau - \tau_0}$, being $\tau_0$ the dimensionless initial time at the moment of impact. Here $R_{Base}$ denotes 
the radius of the crown at its base, as indicated in Fig.~\ref{fig:Intro1}a. This implicitly means that the characteristic time for crown propagation 
is constant during a splashing event and equal to $t_{ref}=D_0/U_0$. This value is indirectly included in the non-dimensional square root 
time dependence through the definition of  $\tau = t/ t_{ref} = U_0 t / D_0$. The constant of proportionality $C_1 = (2/3\delta)^{1/4}$ was 
determined empirically by fitting the experiments of Levin and Hobbs \cite{Levin1971} to describe the effect of wall-film inertia on the rate 
of crown propagation. Over the years, several corrections have been proposed to improve the agreement with experiments by modifying 
the value of the constant $C$ with varying Weber and Reynolds numbers \cite{Cossali2004,Rieber1999,Trujillo2001,Davidson2002,Guo2014,Agbaglah2014,Fujimoto2001,Philippi2016}. A good review 
on the different correlations can be found in \cite{Liang2016}.  Recently, Gao and Li \cite{Gao2015} proposed a unified framework to 
integrate and compare competing modelling approaches. The authors introduced the factor $\lambda = u_{\infty} /U_0$ in the definition of the 
constant $C =  (2 \lambda^2 /3\delta)^{1/4}$. The factor $\lambda$ takes into account momentum losses at the moment of impact due to 
droplet deformation, viscous and inertial forces. This energy loss has two implications. First, it reduces the transfer of tangential momentum 
from the impacting droplet to the crown (kinematic discontinuity) to the potential flow value of $u_{\infty} = \lambda U_0$.  Second, the 
characteristic time for crown propagation is now defined as $t_{ref}=D_0/(\lambda U_0)$. Please note that, in order to maintain the same 
definition of the dimensionless time $\tau$, $\lambda$ was included in the definition of the constant $C$. An empirical correlation was 
proposed for estimating the initial impact losses, according to $\lambda = 0.26 Re_0^{0.05} / (We_0^{0.07} \delta^{0.34})$. The physical 
insight of the Gao and Li's model \cite{Gao2015} is that when the combined effect of deformation, viscous and inertial forces leads 
to the same value of $\lambda$, only then the temporal evolution of the crown base radius will follow the same curve. Independent variation 
of a single parameter (e.g. $\delta, \, Re_0, \, We_0$) will inevitably result in a different value of the constant $C$, thus explaining the 
different correlations proposed in literature. Hereafter, we show how the Gao and Li's approach  \cite{Gao2015} is capable to reconcile 
the predictions from different models, even in the limiting case of impact on a dry wall.

\begin{figure}[t]
\centering
{\includegraphics[width=0.7\columnwidth]{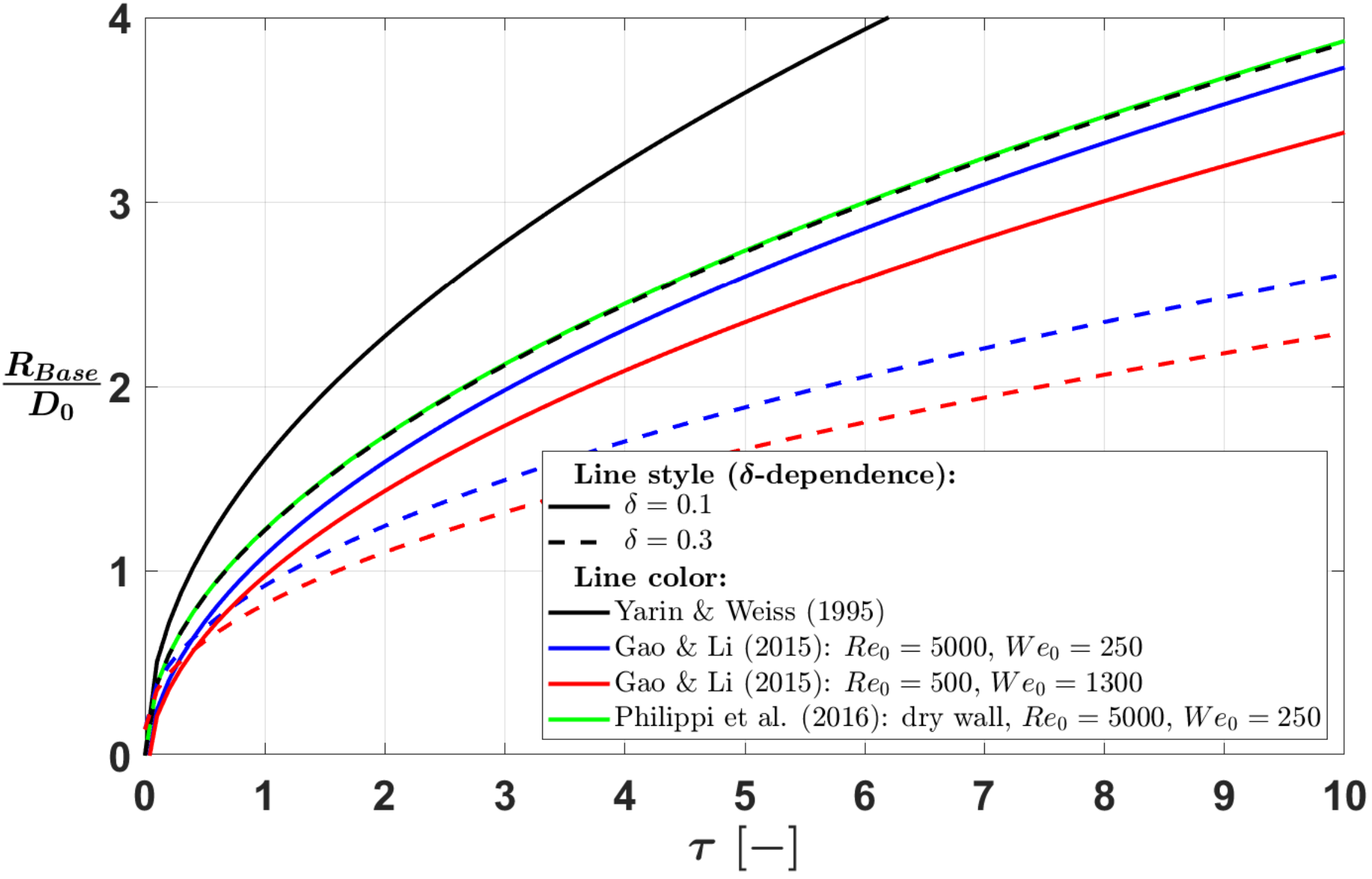}} 
\caption{Temporal evolution of the crown base diameter, as predicted by various theoretical models.} 
\label{fig:oldmod}
\end{figure}

Figure \ref{fig:oldmod} shows a comparison among different theoretical models. The model of Yarin and Weiss~\cite{Yarin1995} includes 
only the effects of wall-film inertia. Hence, it is strictly applicable when viscous and deformation losses are negligible at the moment of 
impact. Indeed, the Yarin and Weiss model was corroborated by the numerical simulations in Refs.~\cite{Rieber1999,Liang2014}, performed 
at high Reynolds [$Re_0 = \mathcal{O}(10^4)$] and low Weber numbers [$We_0 = \mathcal{O}(10^2)$]. Philippi et al.~\cite{Philippi2016}, 
instead, performed numerical simulations of single droplet impact on a dry wall at $Re_0 = 5000$ and  $We_0 = 250$. For the early dynamics 
(up to max $\tau = 1$), they found that the ejecta sheet obeys the following functional dependency $R_{Base}/D_0 = \sqrt{3 U_0 D_0 t/ 
(2 D^2_0)} = \sqrt{3 \tau /2}$, in agreement with the analytical solution of Wagner~\cite{Wagner1932} for describing the water entry of a 
solid body. This conclusion is shared by Riboux and Gordillo~\cite{Riboux2014}, who also applied the potential flow theory of Wagner \cite{Wagner1932} to describe the temporal evolution of the ejecta sheet. Contrary to all expectations, the Philippi et al.'s curve ($\delta = 0$) 
overlaps the Yarin and Weiss' curve for $\delta = 0.3$, while differing significantly from the  $\delta = 0.1$ curve, even in the early dynamics 
of the ejecta sheet (see Fig.~\ref{fig:oldmod}). To explain this apparent discrepancy, the predictions from the Gao and Li's model 
\cite{Gao2015} for the same impact conditions ($Re_0 = 5000$, $We_0 = 250$ and $\delta = 0.1$) are also included in the figure. Thanks 
to the correct inclusion of impact losses, their curve evolves parallel to the Philippi et al.'s curve and overlaps with it for $\delta = 0.03$ in 
the early dynamics. Deviations occur either for increasing $\delta$ at constant $Re_0, \, We_0$ or for increasing viscous (i.e.~lower $Re_0$) 
and deformation losses (i.e.~higher $We_0$) at constant $\delta$. Finally, the Gao and Li's model \cite{Gao2015} was validated over a wide 
range of impact conditions till approximately $\tau = 3$. Due to its accuracy and generalised applicability, it is therefore adopted here as base 
model to describe the propagation of the crown.

\begin{figure}[!hb]             
\vspace{2mm}
    \centering
		\begin{minipage}[b]{0.5\linewidth}
		  \centering
 			{\includegraphics[width=0.95\linewidth]{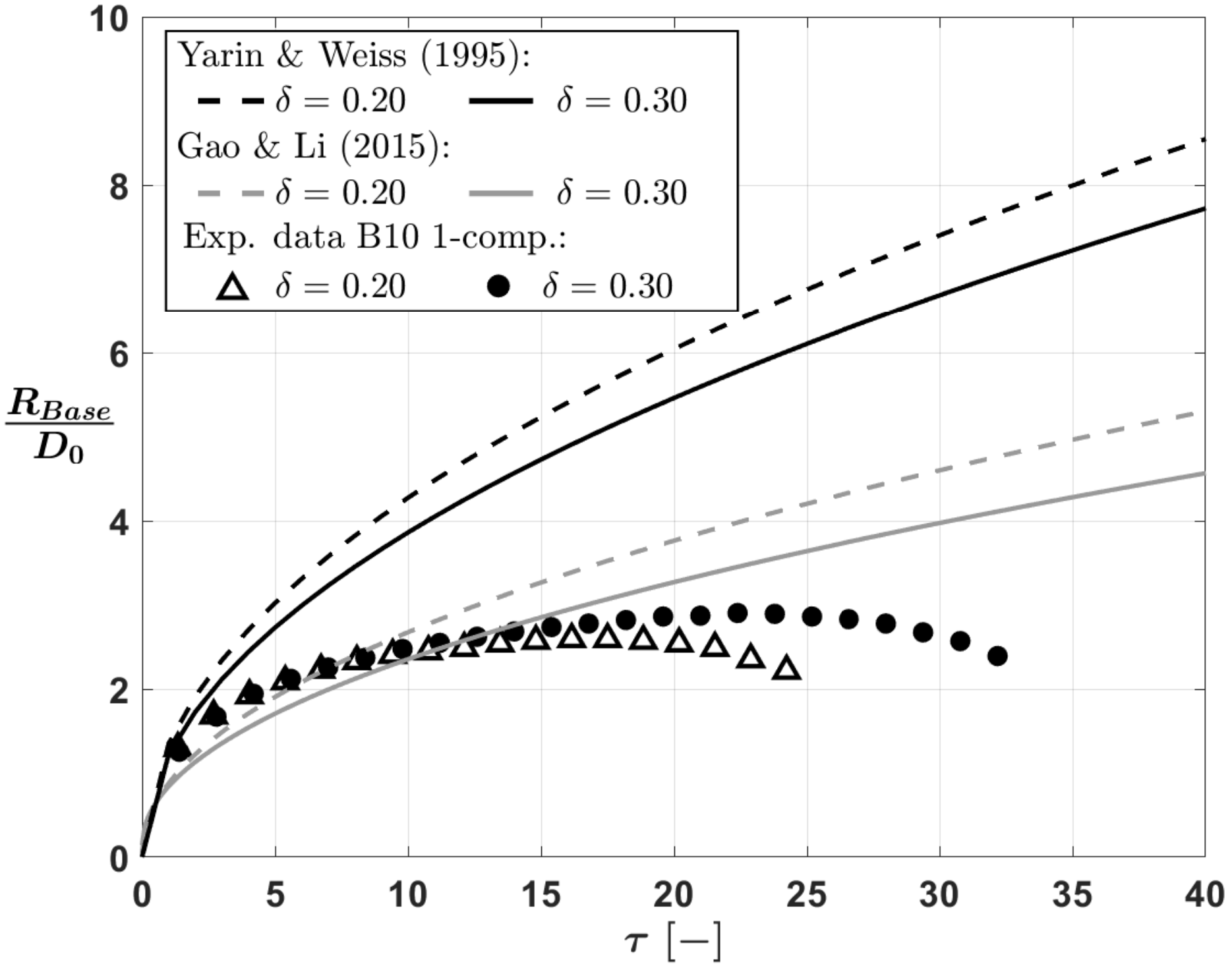}}
			\vspace{2mm}
		  \subcaption{silicone oil B10, $Re_0 = \mathcal{O}(700)$}
   \end{minipage}%
   \begin{minipage}[b]{0.5\linewidth}
			\centering
			{\includegraphics[width=0.95\linewidth]{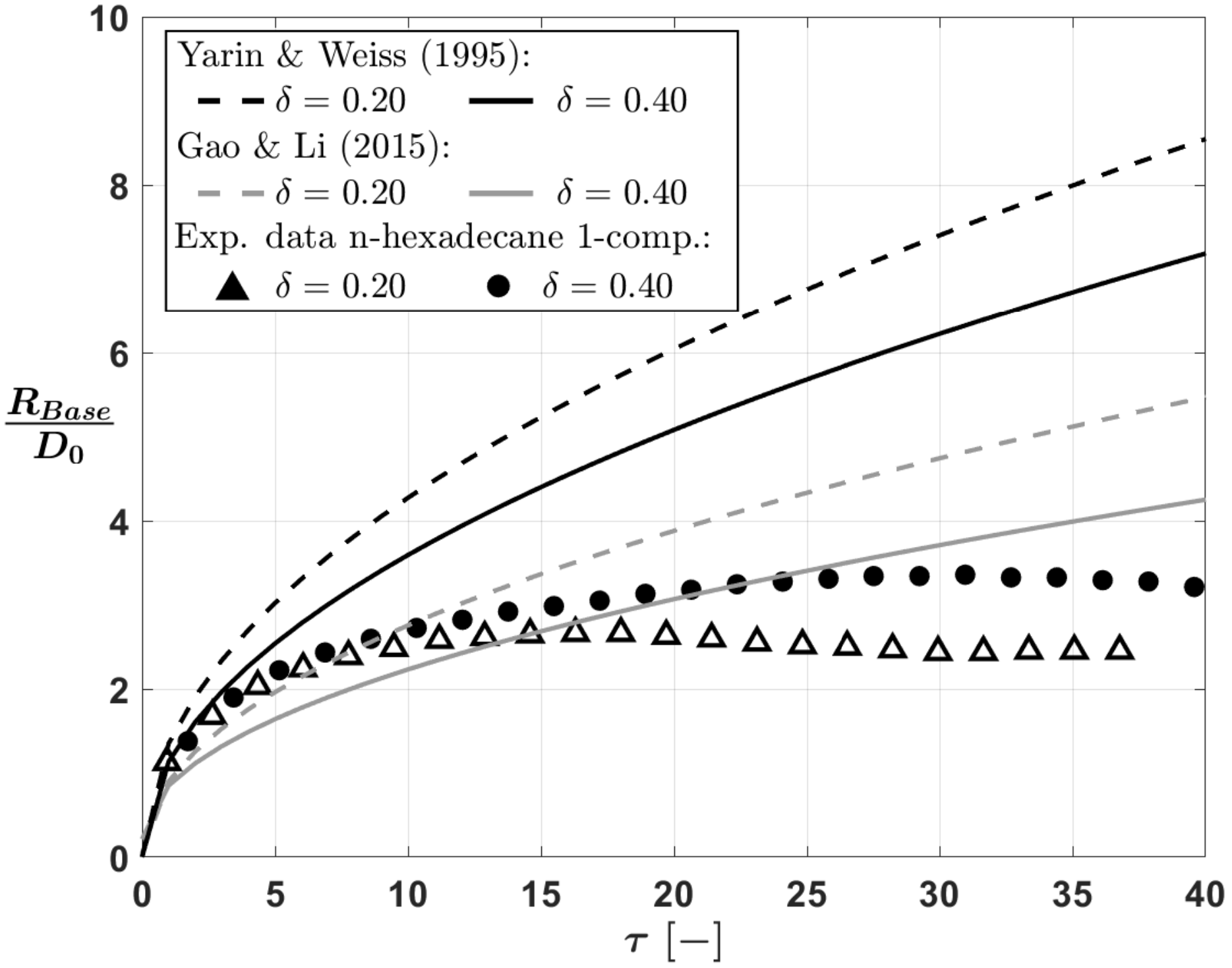}}
			\vspace{2mm}
			\subcaption{\textit{n}-hexadecane, $Re_0 = \mathcal{O}(2400)$}
   \end{minipage}
  \caption{Temporal evolution of the crown base radius: comparison between experiments and theoretical models: (a) Test fluid: silicone 
  oil B10. Impact conditions: $We_0 = \mathcal{O}(1200)$ $Re_0 = \mathcal{O}(700)$. (b)  Test fluid: \textit{n}-hexadecane. Impact 
  conditions: $We_0 = \mathcal{O}(1300)$ $Re_0 = \mathcal{O}(2400)$. The exact test conditions are listed in Table \ref{tab:Exp} (see 
  Appendix \ref{app:TC}).}    
  \label{fig:Intro2}
\vspace{2mm}
\end{figure}

As shown in Fig.~\ref{fig:Intro1}, however, crown bottom breakup occurs around $\tau \approx 7$. It is, therefore, important to assess the 
accuracy of the Gao and Li's model \cite{Gao2015} in describing the crown propagation at later times, particularly with reference to the 
speed of crown propagation that controls the mass flow rate entering the crown lamella. Figure \ref{fig:Intro2} compares the models' predictions 
of the crown base radius $R_{Base}$ with experimental data for two different test fluids. Note that the comparison focuses mainly beyond the 
early stage of the crown propagation (i.e.~for $\tau \geq 2.5$), where the model of Gao and Li \cite{Gao2015} has not been validated yet.
The importance of including impact losses can be inferred by comparing Figs.~\ref{fig:Intro2}a and \ref{fig:Intro2}b, where it is 
clear that the predictions of the Gao and Li model \cite{Gao2015} are, on average, closest to the experimental data, particularly for low 
initial Reynolds numbers (e.g.~$Re_0 = 700$ for B10). However, for both test fluids, the experimental crown spreading rate is only 
weakly dependent upon $\delta$ for $\tau \geq 2.5$, contrary to the predictions of theoretical models. A possible explanation for the 
observed experimental trends is proposed hereafter. In the early stage of crown dynamics ($\delta < 3$), wall-film inertia 
is the dominant physical process that limits the acceleration imparted by the impinging droplet to the quiescent wall film. As a result, 
the rate of crown propagation increases with decreasing film height ($h_0$) in agreement with all previous theories \cite{Cossali2004,Rieber1999,Trujillo2001,Davidson2002,Guo2014,Agbaglah2014,Fujimoto2001,Gao2015} In this early stage, the 
additional influence of viscous and deformation losses is well represented by $Re_0, \, We_0$, based on the characteristic physical 
parameters of the impacting droplet \cite{Gao2015}. In the later stage of crown dynamics ($\delta \geq 3$), viscous losses generated 
during the spreading phase of the lamella become increasingly important, leading to a significant decrease in the crown speed especially 
for small initial film thicknesses. Over time, this effect counteracts the increase in spreading rate associated to the diminished liquid film 
inertia. As a result, the crown spreading rate becomes only weakly dependent upon the initial wall-film thickness $\delta$ in the later stages  
of crown dynamics.

These considerations motivate the present work, where we propose an alternative modelling approach that incorporates explicitly momentum 
losses in the temporal evolution of the crown base radius $R_{Base}$. As a result, the parameter $C$ is no longer constant, but decays in time 
with the decrease in spreading velocity of the lamella base due to viscous losses. This requires an accurate estimation of the strain rate in the boundary layer, which depends not only upon the fluid viscosity, but also upon the thinning rate of the initial wall film. The advantages of this 
approach are twofold. First, it allows for an accurate prediction of the liquid spreading rate during the entire crown elevation phase by enabling 
a smooth transition from the inertia-driven (inviscid theories) to the viscous-controlled regime of crown propagation. Second, it paves the way 
for predicting the inception of crown bottom breakup (CBB) by enabling an accurate estimation of the flow parameters within the spreading 
lamella. Finally, the limitations of the model are discussed in section \ref{sec:limit}.

\section{\label{sec:Model}Modelling Approach} 
\begin{figure}[t]
\centering
{\includegraphics[width=0.7\columnwidth]{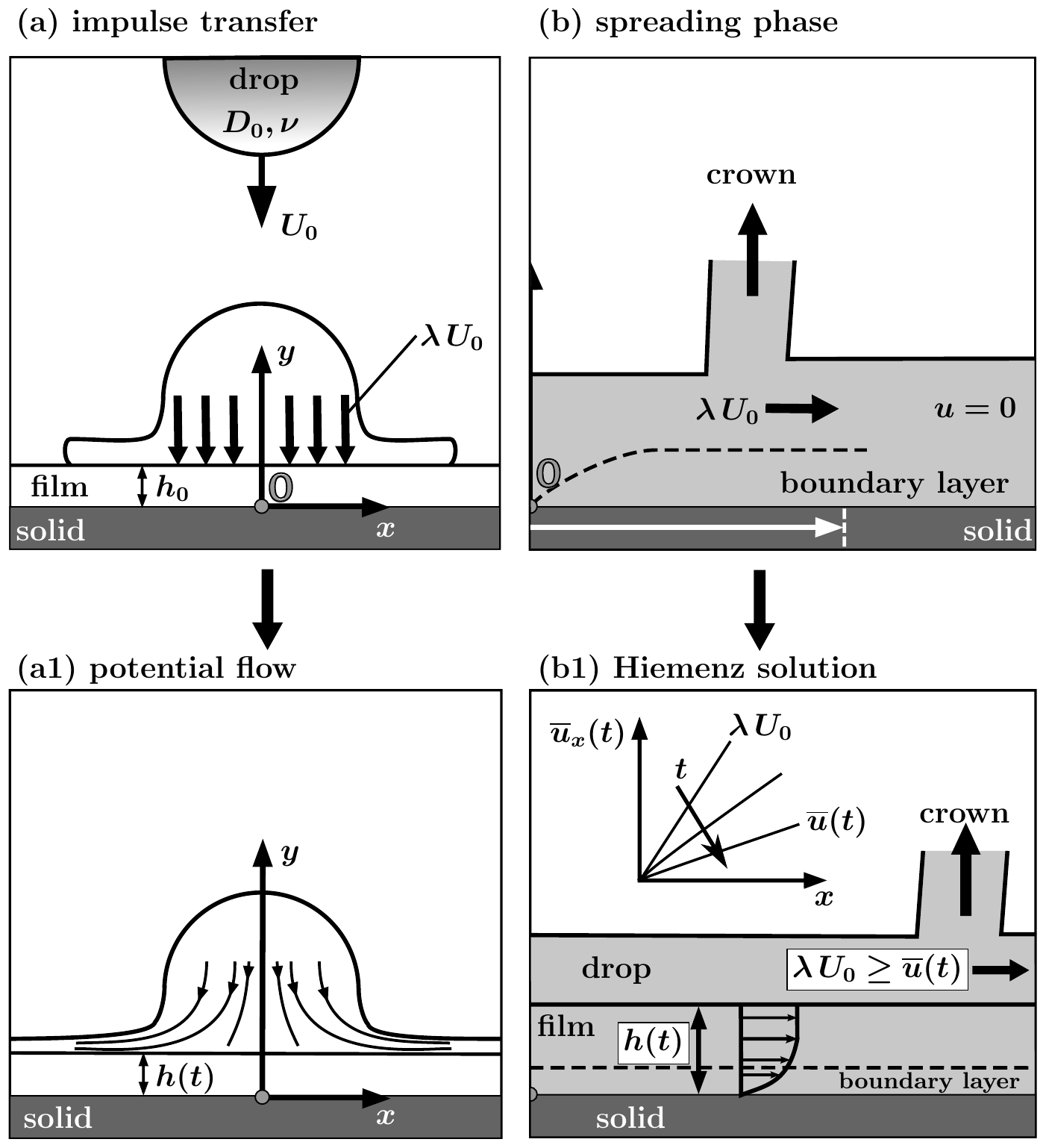}} 
\caption{Scheme of the proposed modelling approach. Phase (a): impulse transfer to the wall film from the impacting drop. The decrease in 
film height is modelled with potential flow theory. Phase (b): radial spreading along the wall with profile-averaged velocity $\bar{u}_x(t)$, which 
decreases progressively in time due to boundary layer effects. Hiemenz's solution is employed to calculate momentum losses.}
\label{fig:modScheme}
\end{figure}

The starting point of our analytical model is the two-dimensional orthogonal stagnation-point flow, due to its geometrical resemblance to the 
droplet impact problem. This approach offers two main advantages. First, it enables a simple estimation of the wall-film thinning rate. Second, 
thanks to the self-similarity of the solution, it allows for a straightforward estimation of the momentum losses. For this purpose, the droplet 
impact on a wet substrate is schematically divided into two sub-processes, as shown in Fig.~\ref{fig:modScheme}. As a result of the collision, momentum is transferred to the wall film (phase a), thereby causing a decrease in wall-film thickness with time $h(t)$. Note that a similar 
assumption is also made by Gao and Li \cite{Gao2015}. Phase (a) is modelled by assuming that, at the moment of impact and in a region 
close to the impact point, the flow inside the droplet resembles the potential flow of a stagnation-point problem. This assumption is corroborated 
by the numerical simulations of Philippi et al.~\cite{Philippi2016} for a droplet collision on a flat rigid surface. Right after the impact, we 
assume that the liquid-droplet and the wall-film fluids merge perfectly and start spreading radially outwards (phase b). The velocity distribution 
along the $x$-axis follows the potential theory. It is zero at the stagnation-point ($x =0$) and increases linearly with $x$ reaching its maximum 
value ($u_{\infty}$) at the crown foot (kinematic discontinuity). Right after the impact, the initial speed of the crown foot is assumed equal 
to $u_{\infty} = \lambda U_0$ to account for impact losses, as suggested by Gao and Li \cite{Gao2015}. Contrary to inviscid models, however, 
the crown propagation dynamics is now directly influenced by boundary layer effects. To estimate momentum losses, the Hiemenz's boundary 
layer solution for a plane stagnation-point flow is employed \cite{Schlichting2017}, because it describes more appropriately the acceleration transferred by the impinging droplet to the liquid wall film. Indicating with $\bar{u}_x(t)$ the line-averaged velocity distribution across the 
spreading lamella, its value will decrease with time due to viscous losses, as schematically indicated in the insert of Fig.~\ref{fig:modScheme}b1. 
In this work, we are mainly interested in estimating the line-averaged velocity of the crown foot (kinematic discontinuity), hereafter denoted 
simply with $\bar{u}(t)$.

\subsection{\label{sec:Potential} Modelling of the wall-film decay} 
This section presents the modelling approach to describe  the decrease in wall-film thickness. Due to the absence of experimental data, we 
closely follow the approach proposed by Blyth and Pozrikidis \cite{Blyth2005}, which was validated against numerical solutions of the steady 
and unsteady Navier-Stokes equations for stagnation-point flow against a wetted plane wall. The origin of the coordinate system is set at the 
point O with the coordinate axis oriented as indicated in Fig.~\ref{fig:modScheme}a. We assume that the flow distribution within the impacting 
droplet is a frictionless potential flow and that sliding effects are negligible at the interface between the impacting drop and the liquid substrate. 
The $x$- and $y$-velocity components of the potential flow can be expressed as $u = a x$ and $v = - a y$. The constant $a$ represents the momentum per unit length transmitted by the droplet to the liquid film at the moment of impact. Its value can be estimated from the initial 
conditions as $a = \lambda U_0 / (\pi D_0)$. Here $\lambda U_0$ is identified as characteristic velocity of the impact strength after removal 
of impact losses, and the circumference $(\pi D_0)$ is chosen as characteristic length scale of the droplet impact footprint.

This interpretation of the impact process is corroborated by the experiments of Mitchell et al.~\cite{Mitchell2019}. The authors measured the 
transient force exerted by a droplet impinging onto a rigid substrate. This transient force exhibits a peak profile followed by an exponential 
decay and induces a vertical displacement of the wall film \cite{Blyth2005,Gao2015}. 
Assuming that the liquid film interface remains always parallel to the wall, its vertical displacement can be evaluated at any $x$ position. 
Following \cite{Blyth2005}, we chose the stagnation-point ($x=0$), where both Stokes theory and potential theory are simultaneously valid.
In the limit of Stokes flow, the $y$-velocity component can be expressed as  $v = - G y^2$. Starting from the kinematic condition at the 
droplet/film interface $y=h(t)$ and being $h_0 = h(0)$ the initial wall-film thickness, it follows that $dh/dt=v=-ay = - G y^2$. For both theories 
to be verified at the moment of impact, it follows necessarily: $a = G h_0$. The physical meaning of the previous relation is the following. 
For a given initial wall-film thickness $h_0$, the stronger the strength of the impacting potential flow, the stronger must be the strength of 
the associated Stokes flow to assure a rapid dissipation of the transferred vertical momentum.

\begin{figure}[!hb]             
\vspace{2mm}
    \centering
		\begin{minipage}[b]{0.5\linewidth}
		  \centering
 			{\includegraphics[width=0.95\linewidth]{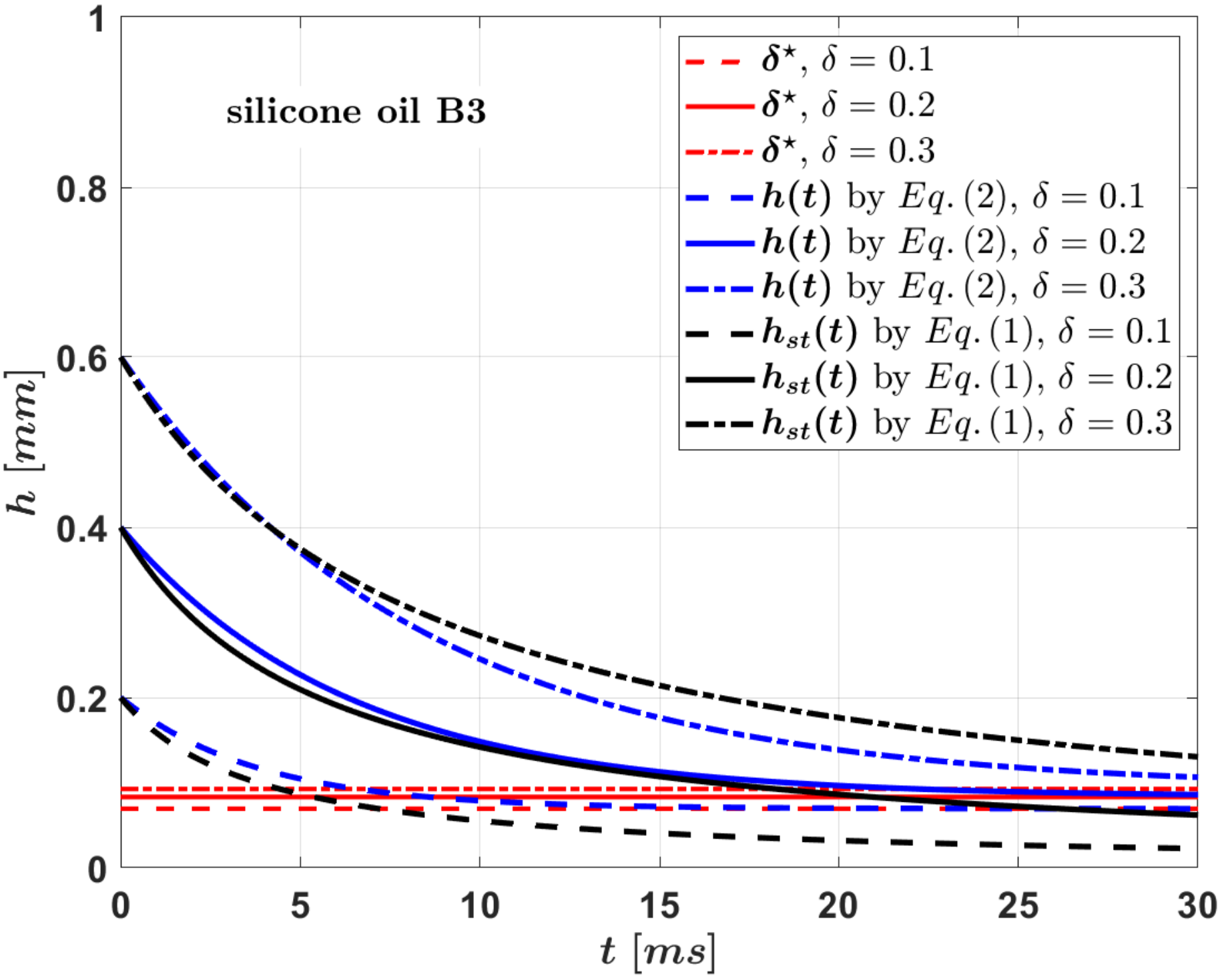}}
			\vspace{2mm}
		  \subcaption{Silicone oil: B3}
   \end{minipage}%
   \begin{minipage}[b]{0.5\linewidth}
			\centering
			{\includegraphics[width=0.95\linewidth]{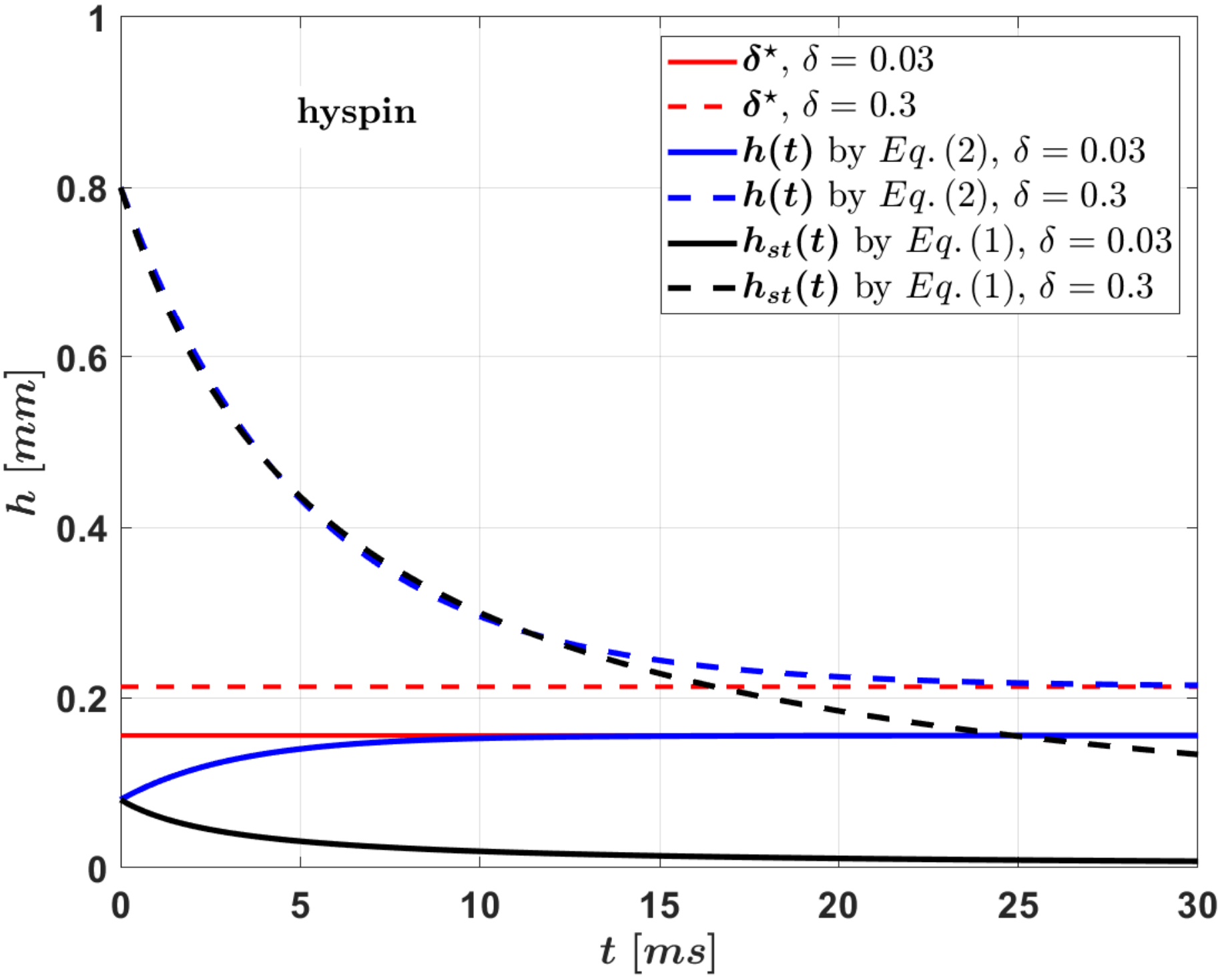}}
			\vspace{2mm}
			\subcaption{Hyspin}
   \end{minipage}
  \caption{Comparison of wall-film decay rate as predicted by Eqs.~(\ref{eq:film}) and (\ref{eq:film2}) for some cases listed in Table \ref{tab:Exp} 
(see Appendix \ref{app:TC}). The two theories are equivalent, except for $\delta^* > h_0$. Then the potential theory is no longer applicable, 
leading to an unphysical increase of $h(t)$, as shown in (b) for the $\delta = 0.03$ case.}
\label{fig:Mod1}
\vspace{2mm}
\end{figure}

In the Stokes flow approximation, integration of the kinematic condition $dh/dt= - G y^2$ between $t = 0$ and $t$ and solving for $h(t)$ 
yields \cite{Blyth2005}
\begin{equation}
h_{st}(t) = \frac{h_0}{1+ G h_0 t} = \frac{h_0}{1+a t}
\label{eq:film}
\end{equation}

\noindent where the subscript ``$st$'' denotes the Stokes solution for the wall-film decay rate. Alternatively, it is also possible to use the potential theory. In order to extend its applicability to viscous fluids, it is necessary to introduce the concept of \emph{displacement thickness} $\delta^*$, 
to include the effects of boundary layer flow on the external potential flow solution. The displacement thickness $\delta^*$ denotes how much 
the wall has to be displaced for an inviscid flow, in  order to have the same mass transport as the viscous flow along the original wall. As shown 
in \cite{Homann1936,Joseph2014,Schlichting2017}, this is equivalent to state that the vertical velocity component of the external flow for a 
viscous fluid must be corrected by a factor $\delta^*$, yielding $v = - a \, (y - \delta^*)$. For a plane stagnation-point flow, the displacement 
thickness is equal to $\delta^* = 0.6479 \sqrt{\nu/a}$ \cite{Homann1936,Schlichting2017}. Hence, integrating the kinematic condition in the 
potential flow approximation $dh/dt= - a \, (y - \delta^*)$ between $t = 0$ and $t$ and solving for $h(t)$ yields 
\begin{equation}
h(t) = \delta^* + (h_0 - \delta^*) exp(-at).
\label{eq:film2}
\end{equation}

Equation (\ref{eq:film2}) shows immediately that the correction to potential theory to account for the fluid's viscosity results in a smoother thinning 
rate of the wall film, since $h(t)$ tends to the limiting value  $\delta^*$. Viscous losses within the drop reduce the momentum transfer to the wall 
film, thereby hampering the decay in wall-film thickness. Similar results have been obtained also by Blyth and Pozrikidis \cite{Blyth2005}. Figure \ref{fig:Mod1} shows the decay of the wall-film thickness for selected experiments from Table \ref{tab:Exp} (see Appendix \ref{app:TC}), 
as predicted by Eqs.~(\ref{eq:film}) and (\ref{eq:film2}). The horizontal lines represent the displacement thickness of the boundary layer. As can 
be seen, both relations predict a very similar decay rate for the wall film, particularly in the relevant time range of splashing events (up to 
roughly $t \approx 25$ ms). The only noteworthy exception occurs when the initial wall-film height $h_0$ is smaller than the displacement thickness $\delta^*$. In this case (shown in Fig.~\ref{fig:Mod1}b), potential theory is no longer applicable and one should revert to the Stokes 
flow approximation. Unless explicitly stated, in this paper we employ the potential flow approximation (i.e.~Eq.~(\ref{eq:film2})) for modelling 
the wall-film vertical displacement.

\subsection{\label{sec:Hiemenz}Estimation of momentum losses} 
Following the scheme proposed in Fig.~\ref{fig:modScheme}b, it is clear that friction losses will lead to an additional decrease in the speed of 
crown propagation (kinematic discontinuity), compared to the initial value ($\lambda U_0$) that accounts only for impact losses. This additional 
decrease will be more and more significant as the wall-film height reaches the boundary layer thickness. For estimating the momentum losses, 
it is necessary to determine the velocity profile within the boundary layer of a stagnation-point flow (SPF). Following Hiemenz's approach \cite{Schlichting2017}, the Navier-Stokes equations can be reduced to an ordinary differential equation by the following 
transformation (plane SPF):
\begin{equation}
\eta = \sqrt{\frac{a}{\nu}}y  \quad  \quad f(\eta) = \frac{\psi}{x \sqrt{\nu a}}
\label{eq:Hiemz1}
\end{equation}
where $\psi$ is the stream function. In the stagnation-point region, the velocity components can be then expressed as $u = u_{\infty}(x) f'(\eta) 
= a x f'(\eta)$ and $v = - \sqrt{\nu a} f(\eta)$. The boundary layer velocity profiles tend asymptotically to the outer potential flow solution. The transformed ordinary 
differential equation reads then as follows:
\begin{equation}
f''' + f f'' +1 -f'^2 = 0
\label{eq:Hiemz2}
\end{equation}
subject to the boundary conditions of no-slip and inviscid flow limit:
\begin{equation}
\eta = 0: \quad f = 0, f' = 0; \quad \quad \eta \rightarrow \infty: \quad f' = 1.
\label{eq:HiemzBQ}
\end{equation}

Note that self-similar solutions can be also obtained for a steady and unsteady axisymmetric stagnation-point flow (SPF) \cite{Drazin2007}. As 
shown in \cite{Schlichting2017}, the steady axisymmetric SPF solution differs only slightly from the planar solution and therefore it is not 
considered here. The importance of unsteady effects is discussed in Appendix \ref{App:unsteady}, where it is shown that, for values of the 
parameter $a \geq 100$, unsteady effects can be safely neglected in the modelling of momentum losses during the spreading of a liquid 
lamella. This statement is corroborated by recent theoretical and numerical findings. Riboux and Gordillo \cite{Riboux2017} modelled droplet 
impact on a dry wall as a boundary layer flow with an outer (stagnation-point) potential flow. By performing an analysis of the order of magnitude 
between the local and the convective acceleration terms in the momentum equation, they concluded that the unsteady term can be neglected 
in the boundary layer equations and obtained for the non-dimensional stream function $f(\eta)$ a transport equation identical to Eq.~(\ref{eq:Hiemz2}). Philippi et al.~\cite{Philippi2016} computed numerically the unsteady Navier-Stokes flow for an impacting droplet and 
found that the boundary layer velocity profiles at different radial locations exhibit a self-similar behaviour. The numerical profiles display only 
a weak radial dependence to the point that, when compared to the Blasius' solution for a steady boundary layer along a flat plate, only slight 
deviations between theory and numerical results were observed. 

\begin{figure}[!ht]
\centering
{\includegraphics[width=0.6\columnwidth]{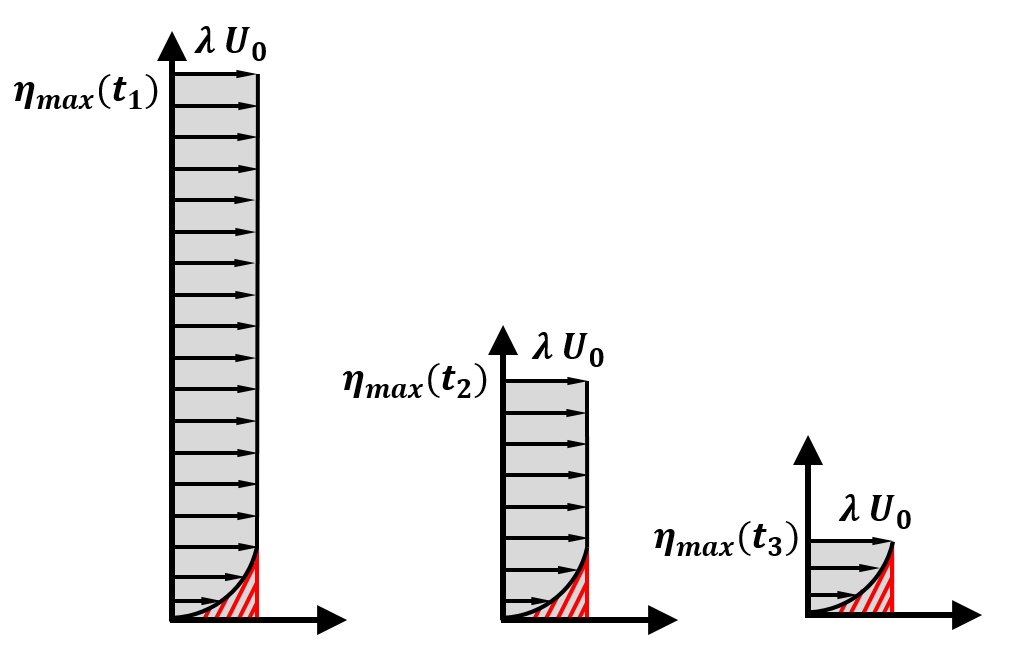}} 
\caption{Graphic visualisation of the proposed modelling approach for estimating momentum losses. The grey vector area represents  
the momentum carried by the lamella, deprived of momentum losses (red striped area). Equation (\ref{eq:Hiemz3b}) represents the 
associated profile-averaged velocity.}
\label{fig:integral}
\end{figure}

Based on these findings, we will therefore use the steady and planar stagnation-point flow solution for the estimation of momentum losses 
in the boundary layer. The latter can be estimated by introducing a profile-averaged non-dimensional velocity according to:
\begin{equation}
\bar{f}' = \frac{\bar{u}}{u_{\infty}} = \frac{1}{\eta_{max}} \int_{0}^{\eta_{max}} f' d\eta
\label{eq:Hiemz3b}
\end{equation}

where $\eta_{max}$ represents the scaled height of the wall film, i.e.~$\eta_{max} = \sqrt{a/ \nu} \; h(t)$. This modelling assumption 
automatically implies that frictional losses are confined in the wall-film flow, while at the interface the inviscid flow velocity is recovered. This assumption is hereafter denoted as ``no sliding'' assumption, as shown in Fig.~\ref{fig:BLscheme} where the limitations of the model are 
discussed. Equation (\ref{eq:Hiemz3b}) essentially estimates the momentum carried by the moving lamella by calculating the total area of 
the velocity profile and dividing it by the non-dimensional film height. Its physical meaning is schematically illustrated in Fig.~\ref{fig:integral}. 
The momentum losses are represented by the red area. In the inviscid approximation (see e.g.~\cite{Gao2015}), the velocity is constant 
along the wall film (grey rectangular area). Hence, Eq.~(\ref{eq:Hiemz3b}) yields $\bar{u}=u_{\infty} = \lambda U_0 $. If we now incorporate 
a boundary layer flow close to the wall, Eq.~(\ref{eq:Hiemz3b}) measures the average momentum carried by the lamella $\bar{u}$, deprived 
of momentum losses, as a fraction of $u_{\infty}$.
The integration of Eq.~(\ref{eq:Hiemz3b}) yields no integration constant, since both functions $f'(\eta)$ and $f(\eta)$ cross the 
origin. Hence, the non-dimensional profile-averaged velocity can be expressed as
\begin{equation}
\bar{f}' = \frac{\bar{u}}{u_{\infty}} = \frac{1}{\eta_{max}} f(\eta_{max}).
\label{eq:Hiemz4}
\end{equation}
Note that $\bar{f}'$ is not constant in time, as it is implicitly a function of the film height $h(t)$. Hence, even though the self-similar solution 
$f(\eta)$ does not change in time, the profile-averaged non-dimensional velocity $\bar{f}'$ will vary between approximately one (when  
$h(t) \gg h_{BL}$) and zero (when $h(t) \rightarrow 0$). The temporal evolution of $\bar{f}'$ and its dependence upon the initial wall-film 
thickness is  discussed in section \ref{sec:Result} for a few representative cases. The analytical solution $\bar{f}'$ can now be employed to 
estimate the spreading velocity $\bar{u}(t)$ of the crown.  As a first step, the averaged velocity is transformed back to the physical coordinate 
system according to
\begin{equation}
\frac{\bar{u}(t)}{u_{\infty}} = \lambda_1(t) = \frac{1}{\sqrt{\frac{a}{\nu}} h(t)} f  \left( \sqrt{\frac{a}{\nu}} h(t) \right).
\label{eq:Hiemz5}
\end{equation}
Here $u_{\infty}$ denotes the velocity outside the boundary layer as determined by potential theory and $\lambda_1(t)$ represents 
the loss in momentum due to viscous forces. In the present work, we are only interested in determining the velocity of propagation of the 
crown (kinematic discontinuity). As stated earlier, we follow here the approach of Gao and Li \cite{Gao2015}. Therefore, for the spreading phase, 
we assume for the ejecta sheet $u_{\infty} = \lambda U_0 = const.$ to take into account the energy losses occurring during droplet impact. Essentially, Eq.~(\ref{eq:Hiemz5}) simply states that the time-dependent, profile-averaged crown speed $\bar{u}(t)$ is a fraction of the initial transmitted velocity $\lambda U_0$. The value $\bar{u}(t)$ can now be incorporated in the modelling of the crown base radius $R_{Base}$. 
Following Gao and Li \cite{Gao2015}, it is more convenient to scale the crown speed with respect to the initial droplet impact velocity $U_0$, 
in order to maintain the same definition of the dimensionless time $\tau$. Hence, based on Eq.~(\ref{eq:Hiemz5}), we can introduce a modified correction factor $\lambda_{AG}$ according to
\begin{equation}
\lambda_{AG}(t) = \frac{\bar{u}(t)}{U_0} = \frac{\lambda}{\sqrt{\frac{a}{\nu}} h(t)} f \left( \sqrt{\frac{a}{\nu}} h(t) \right).
\label{eq:Hiemz6}
\end{equation}
Hence, the parameter $\lambda_{AG}(t)$ measures the decrease in the crown's propagation speed due to both impact losses 
(i.e.~$\lambda$-dependence) and viscous losses (i.e.~$\lambda_1(t)$-dependence). At this stage, it is important to realise that the above 
definition of $\lambda_{AG}(t)$ has important implications on the characteristic experimental time $t_{ref}$. The latter now increases during 
crown propagation according to $t_{ref}=D_0/\bar{u}(t)=D_0/(\lambda_{AG}(t) U_0)$. This implies that the total duration of a splashing event 
will increase with increasing wall-film thickness $\delta$ and fluid kinematic viscosity. This statement will be verified in section \ref{sec:Result} 
through comparison with experiments. The crown base evolution is then modelled as follows
\begin{equation}
\frac{R_{Base}}{D_0} = 0.5 +  \left ( \frac{2 \lambda^2_{AG}(t)}{3 \delta} \right )^{1/4} \sqrt{\tau}.
\label{eq:Hiemz7}
\end{equation}

Both in the original Gao and Li's model \cite{Gao2015} and in Eq.~(\ref{eq:Hiemz7}), the temporal offset $\tau_0$ is set to zero, because the 
time of impact can be accurately determined in both experiments. They differ, however, in two main aspects. The first fundamental difference 
consists in the use of the variable factor $\lambda_{AG}(t)$ that takes into account not only impact losses, but also the additional decrease of lamella's propagation speed due to viscous losses. The second difference lies in the choice of the vertical intercept. In Gao and Li \cite{Gao2015}, 
the intercept is determined empirically as function of the initial film thickness to take into account that the initial spreading of the crown 
radius is less fast for larger film thicknesses. In Eq.~(\ref{eq:Hiemz7}), a constant value is chosen. This choice, albeit not optimal for all 
experimental conditions, corresponds to the tracking of the crown radius, which starts as soon as the droplet is no longer visible in the 
images. This occurs approximately when the crown radius equals the droplet radius, thus explaining the 0.5 shift. More details on the definition 
and post-processing of the crown base radius can be found in Appendix \ref{app:Exp}. Note, incidentally, that the exact definition of the vertical intercept does not affect the slope of the $R_{Base}$ curve, which is the main focus of the present work.

\section{\label{sec:Result}Results and Discussions}
\begin{figure}[!t]
\centering
{\includegraphics[width=0.7\columnwidth]{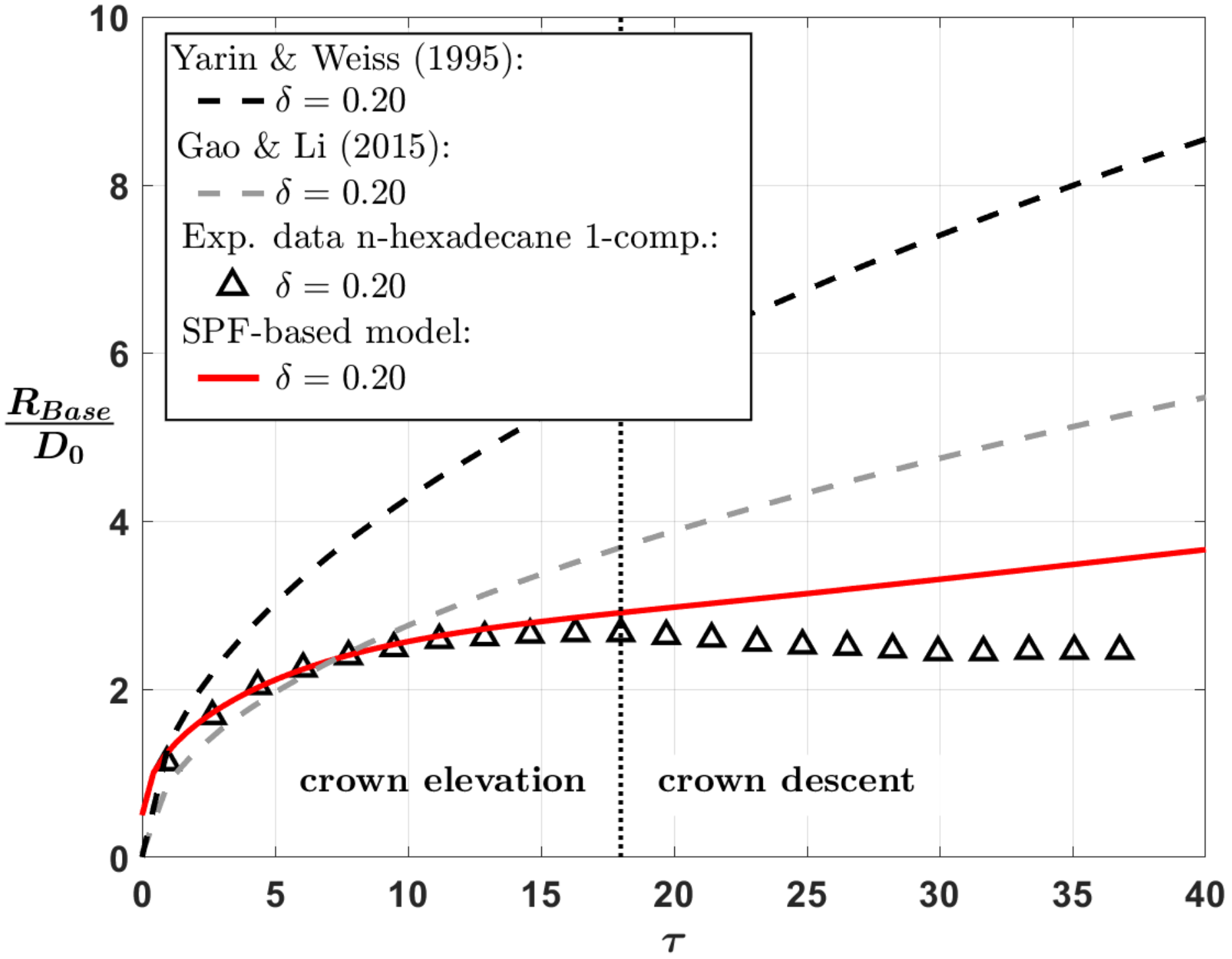}} 
\caption{Experimental and theoretical evolution of the crown base radius $R_{Base}$ for $We=1300$ and $Re=2400$. Fluids: 
\textit{n}-hexadecane for both droplet and wall film. Test conditions are listed in Table \ref{tab:Exp} (see Appendix \ref{app:TC}).}
\label{fig:Res1}
\end{figure}

This section presents a critical analysis of the effectiveness of Eq.~(\ref{eq:Hiemz7}) in reproducing the temporal evolution of the crown base 
radius for a variety of fluids and initial wall-film thicknesses. In the process, both the strength and the limitations of the model are highlighted. 
As a first step, one of the \textit{n}-hexadecane experiments discussed in the introduction (see Fig.~\ref{fig:Intro2}b) is recalculated here with 
the stagnation-point flow model (Eq.~\ref{eq:Hiemz7}) and shown in Fig.~\ref{fig:Res1}. Contrary to inviscid models, our modelling approach 
follows accurately the experimental curve during the entire crown ascending phase ($\tau \approx 18$).
Deviations are instead observed during 
the receding phase, where capillary forces - not considered in the stagnation-point flow (SPF) model - cause the contraction of the crown, as discussed in more details in section \ref{sec:limit}. Once again we explicitly point out that the main difference between the inviscid and the 
SPF-based approach lies in the inclusion of viscous losses during the spreading phase of the lamella.  Yarin and Weiss  \cite{Yarin1995} 
assume the characteristic experimental time $t_{ref}$ to be constant and equal to $t_{ref} = const. = D_0/U_0$, while Gao and Li \cite{Gao2015} assume $t_{ref}  = const. = D_0 /(\lambda U_0)$. The SPF-based model, instead, determines the average momentum losses from the solution 
of the boundary layer flow within the wall film. This leads to an additional decrease in speed of crown propagation $\bar{u}(t) = \lambda_{AG}(t) 
U_0 = \bar{f}'(t) u_{\infty}$ and to a time-dependent characteristic experimental time $t_{ref}(t)  = D_0 /(\lambda_{AG}(t) U_0)$. Here an 
analogy can be drawn with the notion of diffusion length in unsteady diffusion problems. As the diffusion length increases with the square root 
of time at a rate that depends upon the characteristic time for diffusive transport, similarly the axial penetration of the kinematic discontinuity (i.e.~$R_{Base}$) increases in time at a rate that varies with the local characteristic time for convective transport $t_{ref}(t)$. The latter is 
inversely proportional to the speed of crown propagation. Therefore, it is important to evaluate how strong the effects of viscous losses are 
on the velocity decay. 
\begin{figure}[!b]
\centering
{\includegraphics[width=0.7\columnwidth]{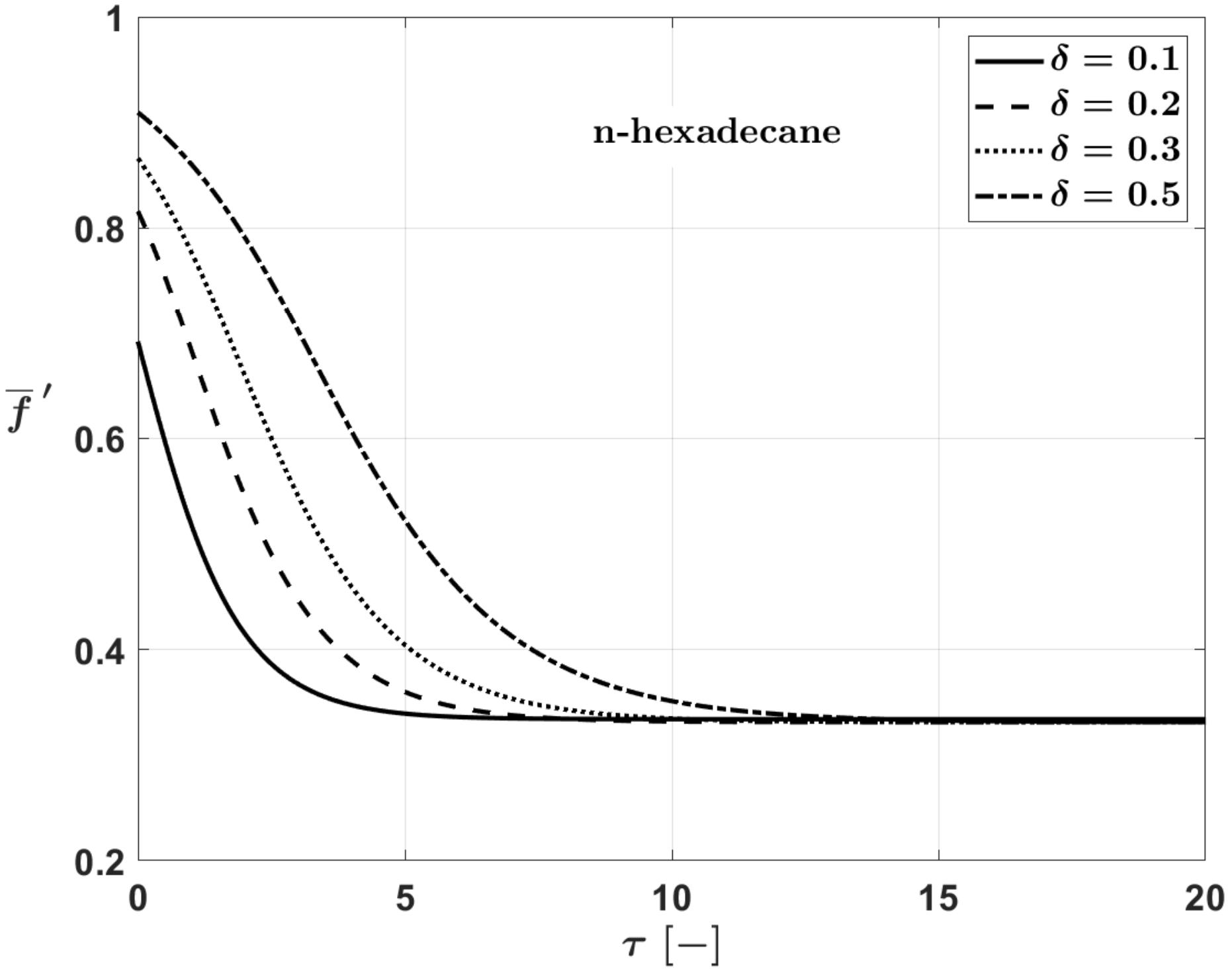}} 
\caption{Temporal evolution of the non-dimensional, profile-averaged crown speed for selected experiments. Test conditions can be found 
in Table \ref{tab:Exp} (see Appendix \ref{app:TC}).}
\label{fig:Mod3}
\end{figure}
Figure \ref{fig:Mod3} shows the non-dimensional, temporal evolution of $\bar{u}(t)$, i.e.~$\bar{u}(\tau)/u_{\infty} = \bar{f}'$, for a number of representative test cases, listed in Table \ref{tab:Exp}. Note that $\bar{u}(t)$ has been normalised with respect to $u_{\infty} = \lambda U_0$, 
in order to single out only the viscous losses generated during the spreading of the lamella along the wall. As can be seen, the latter cannot 
be neglected during a significant portion of crown propagation and induce a significant decrease in crown speed. In particular, the deceleration experienced by the crown becomes more prominent with decreasing wall-film thickness. 
The analytical solution can now be employed to investigate the temporal evolution of the crown velocity $\bar{u}(t)$ and wall-film 
height $h(t)$ with varying fluid properties and initial wall-film thickness $h_0$. Finally, our analytical solutions for $R_{Base}(t)$, $\bar{u}(t)$ 
and $h(t)$ can be employed to estimate the mass flow rate [$q(t) = 2 \pi  R_{Base}(t) \, \rho \, \bar{u}(t) h(t)$], entering the upraising lamella 
during the elevation phase of the crown. As shown in section \ref{sec:CBB}, the associated decrease in mass flow rate $q(t)$ is one of the 
primary factors leading to the destabilisation of the lamella at the onset of crown bottom breakup (CBB). 

\subsection{\label{sec:Viscosity}On the role of viscous losses} 
\begin{figure}[!t]
\centering
{\includegraphics[width=0.7\columnwidth]{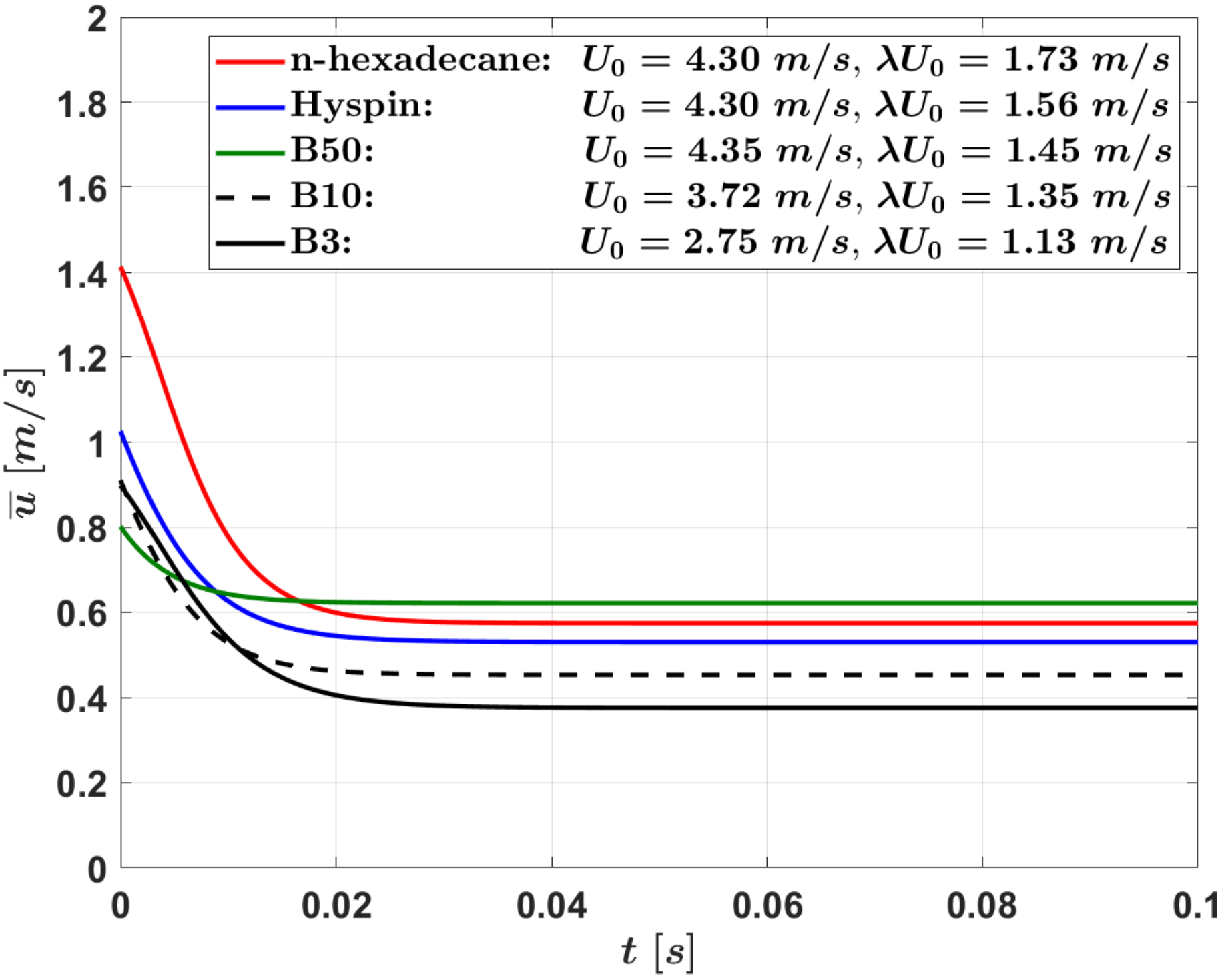}} 
\caption{Temporal evolution of the crown profile-averaged velocities for different test fluids and constant initial film thickness $\delta = 0.2$. 
All other experimental parameters can be found in Table \ref{tab:Exp} (see Appendix \ref{app:TC}). The profile-averaged crown velocity in 
physical units $\bar{u}(t)$ is calculated from Eq.~(\ref{eq:Hiemz6}) as $\bar{u}(t) = \lambda_{AG}(t) U_0$.}
\label{fig:viscosity}
\end{figure}

This section analyses the influence of viscous losses on the propagation of the crown, which depends upon both fluid viscosity and the velocity 
gradient across the wall film. For this purpose, aside of \textit{n}-hexadecane and hyspin, silicone oils are also included as test fluids. This 
enables to vary systematically the fluid viscosity, while keeping all other fluid properties basically constant. As mentioned already in section \ref{sec:Hiemenz}, we explicitly point out that, due to the self-similar behaviour of the boundary layer flow, the analytical solution $f(\eta)$ is unaffected by the specific choice of test fluids or initial wall-film height. The influence of viscosity is retrieved only in Eq.~(\ref{eq:Hiemz5}), 
where the non-dimensional velocity is scaled back to the physical coordinate system. Note that the influence of viscosity becomes noticeable 
in three ways: explicitly in the pre-factor $\sqrt{a / \nu}$, implicitly in the modelling of the wall-film decay $h(t)$ (see Eq.~\ref{eq:film2}) and in 
the empirical correlation for $\lambda$, where it induces higher impact losses. In order to understand how fluid viscosity and $\delta$ affect 
the crown propagation, the temporal evolution of the profile-averaged crown speed is analysed by varying the following parameters: impact 
velocity, initial film thickness and test fluids. Figure \ref{fig:viscosity} shows that, for constant impact velocity $U_0 \approx 4.3$ m/s and film 
thickness $\delta = 0.2$ (see red, blue and green lines), the starting value of $\bar{u}(t)$ decreases with increasing fluid viscosity due to 
the increased momentum losses during droplet impact. As a result, the momentum transferred to the wall film is reduced, leading to lower 
values of the parameter $\lambda \, U_0$. Viscosity, however, is not the only parameter affecting the temporal evolution of $\bar{u}(t)$. 
Indeed, the velocity profiles for the B3 and B10 cases are very similar, despite an increase in fluid viscosity by a factor 3.3. This result is 
attained thanks to the increase of impact velocity $U_0$ for the B10 case and, most important, thanks to the slower decay of $h(t)$ for B10 
(see Eq.~\ref{eq:film2}), which attenuates the retarding effect of the solid substrate.

\begin{figure}
\vspace{2mm}
    \centering
		\begin{minipage}[b]{\linewidth}
		  \centering
 			{\includegraphics[width=0.5\linewidth]{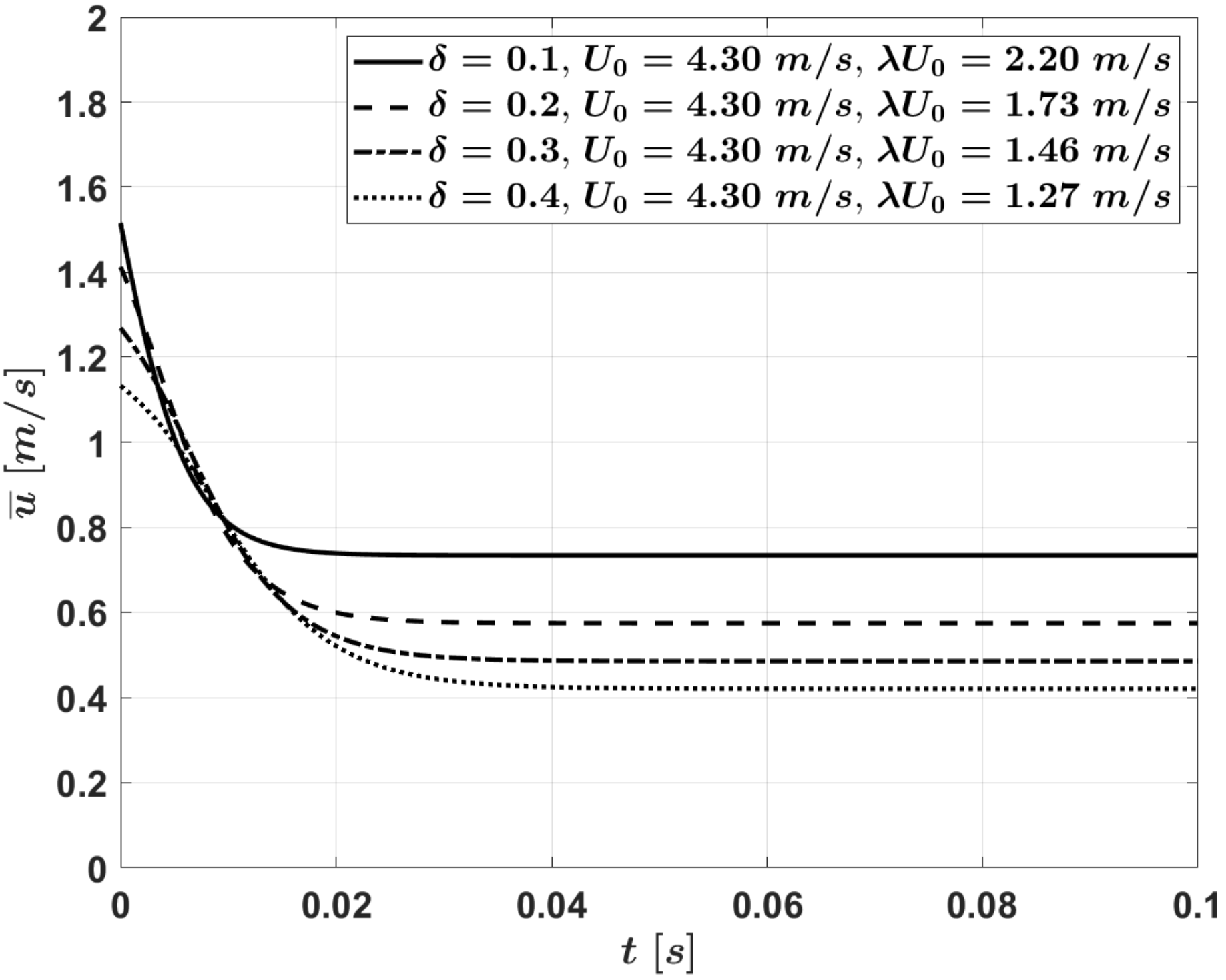}}
			\vspace{2mm}
		  \subcaption{\textit{n}-hexadecane}
   \end{minipage}
   \begin{minipage}[b]{\linewidth}
			\centering
			{\includegraphics[width=0.5\linewidth]{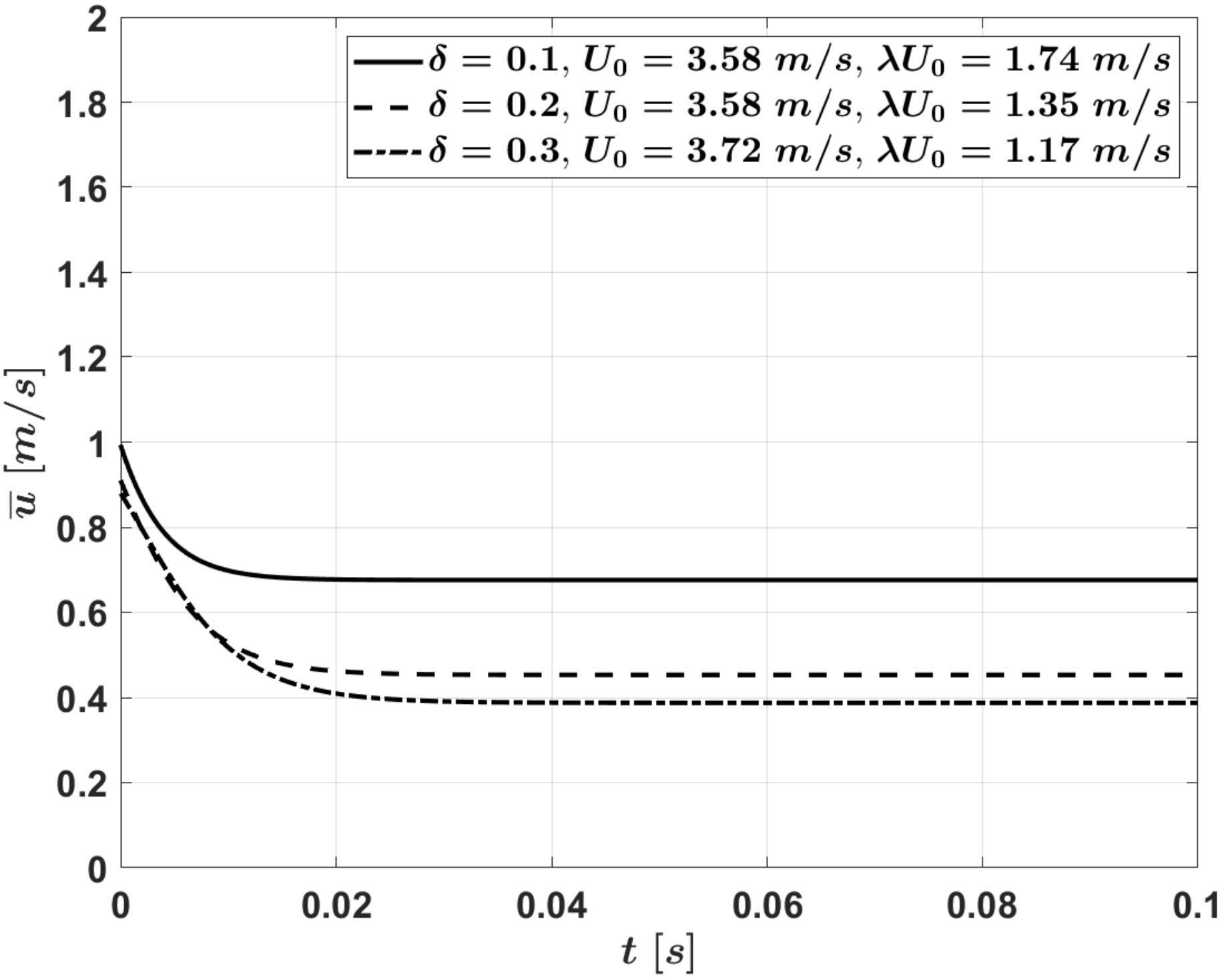}}
			\vspace{2mm}
			\subcaption{Silicon oil: B10}
   \end{minipage}
	   \begin{minipage}[b]{\linewidth}
			\centering
			{\includegraphics[width=0.5\linewidth]{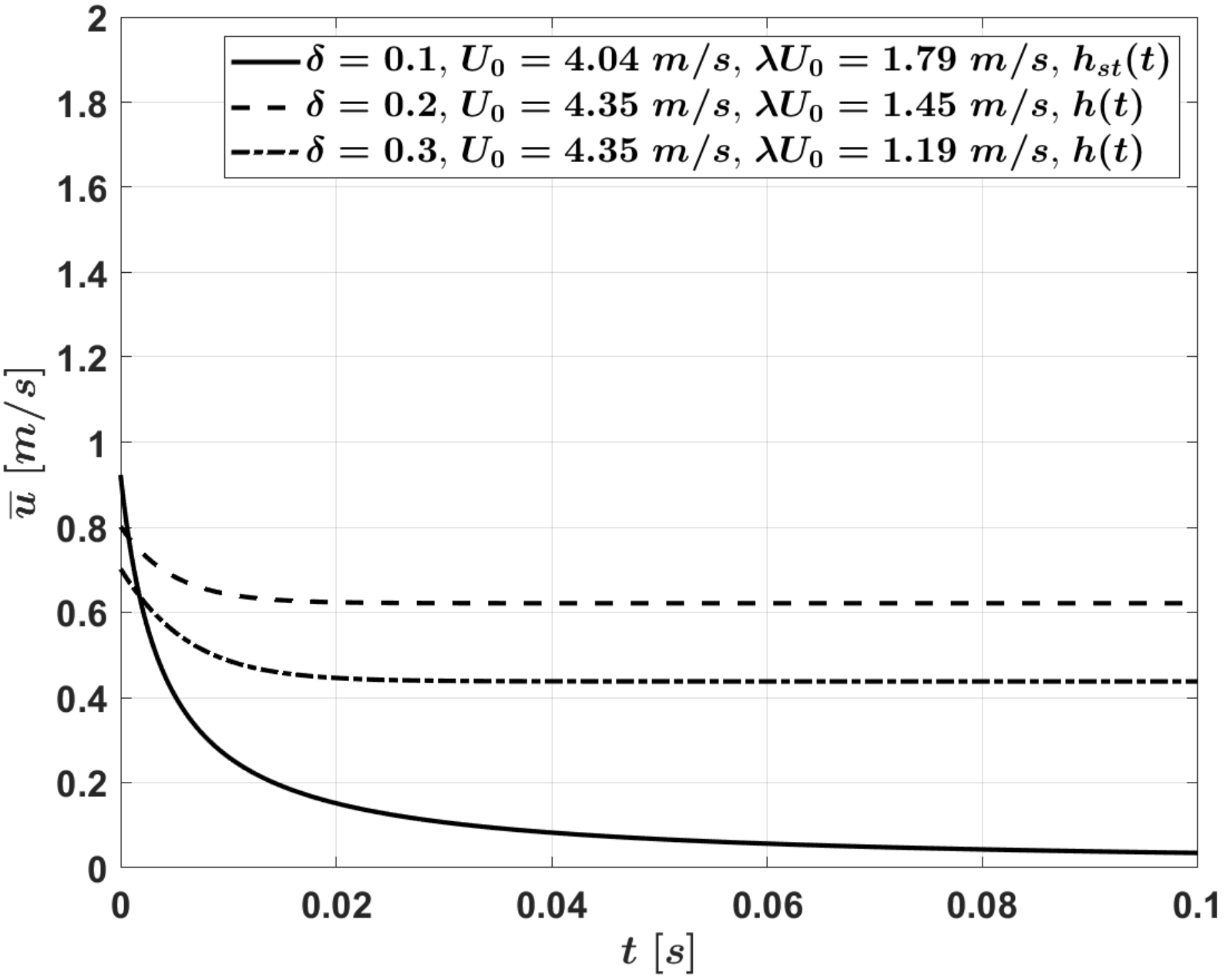}}
			\vspace{2mm}
			\subcaption{Silicon oil: B50}
   \end{minipage}
  \caption{Effect of wall-film inertia on the temporal evolution of the crown profile-averaged velocities for three test fluids.
	Test conditions are listed in Table~\ref{tab:Exp} (see Appendix \ref{app:TC}).}
  \label{fig:inertia}
\vspace{2mm}
\end{figure}

The effect of wall-film inertia is shown in Fig.~\ref{fig:inertia} for three representative test fluids. In general, for constant impact velocity $U_0$, 
a decrease in wall-film thickness $\delta$ will result in lower inertial losses and consequently in an enhanced momentum transfer to the spreading lamella, i.e.~larger potential flow velocity $\lambda U_0$. However, due to the concomitant increase in viscous losses with decreasing $\delta$, 
this excess in initial momentum is rapidly dissipated. As a result, in most cases, only marginal differences are observed in the spreading velocity 
of the lamella towards the end of a splashing experiment, typically in the range of 10-30 ms depending on the initial film thickness. With reference 
to the modelling of the crown base radius $R_{Base}$ (see Eq.~\ref{eq:Hiemz7}), the parameter $C(t) = [2 \lambda^2_{AG}(t) /3 \, \delta]^{1/4}$ experiences two opposite and concurring effects with decreasing $\delta$. Specifically, $C(t)$ increases with decreasing wall-film inertia and simultaneously reduces due to the associated reduction in the parameter $\lambda_{AG}$. Hence, the increase in viscous losses with decreasing 
$\delta$ partially counterbalances the associated decrease of inertial forces. This concomitant interplay between viscous and inertial forces on 
the temporal evolution of the crown base radius ($R_{Base}$) is a key feature of the present model, which is not entailed in any of the inviscid models. It also provides an explanation for the experimental findings in Fig.~\ref{fig:Intro2}, where the crown base radius evolution was found 
to be independent of $\delta$ for a large part of the crown dynamics.
\begin{figure}
\vspace{2mm}
    \centering
		\begin{minipage}[b]{0.5\linewidth}
		  \centering
 			{\includegraphics[width=1\linewidth]{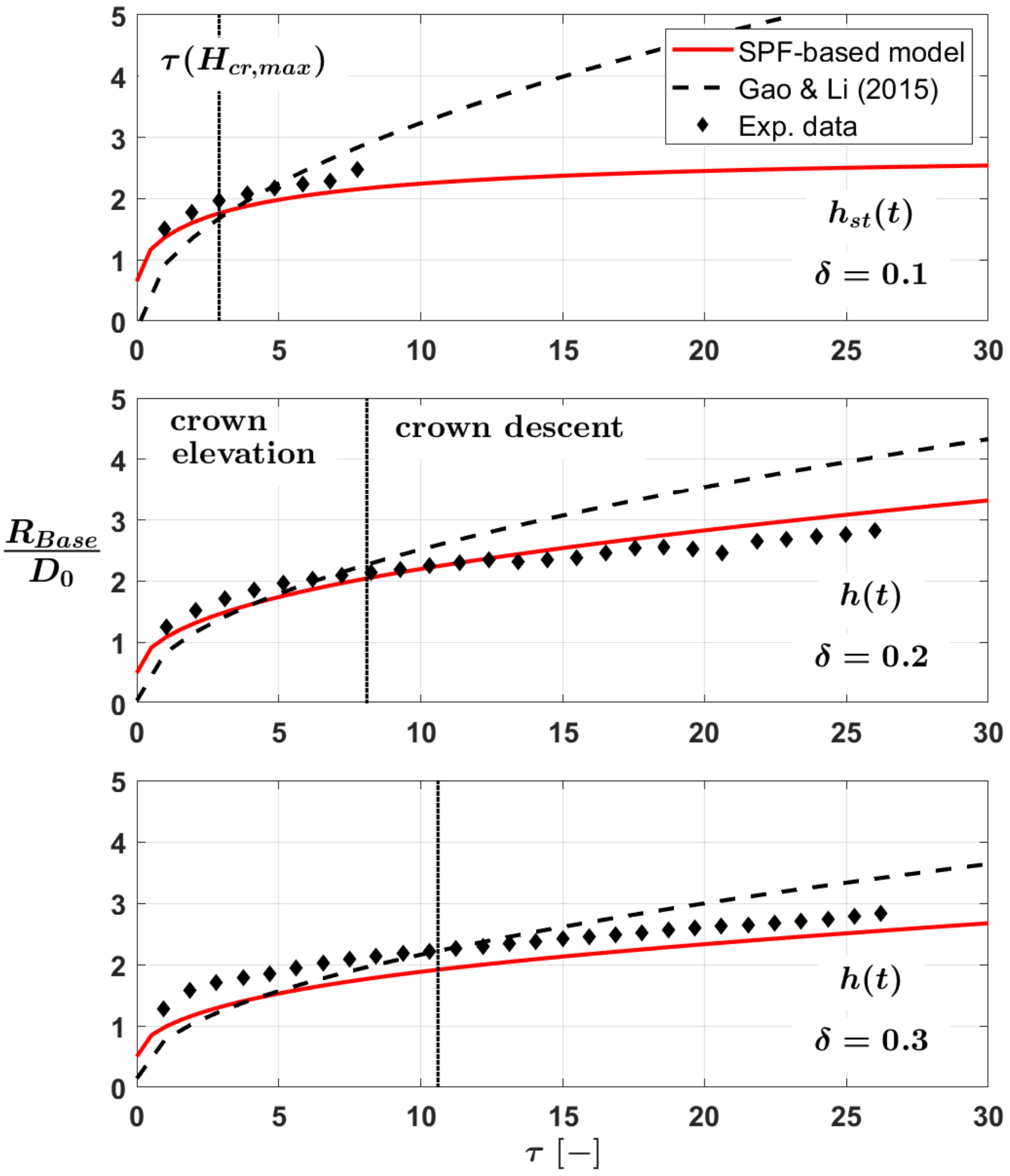}}
		  \subcaption{Silicon oil: B50}
   \end{minipage}%
   \begin{minipage}[b]{0.5\linewidth}
			\centering
			{\includegraphics[width=1\linewidth]{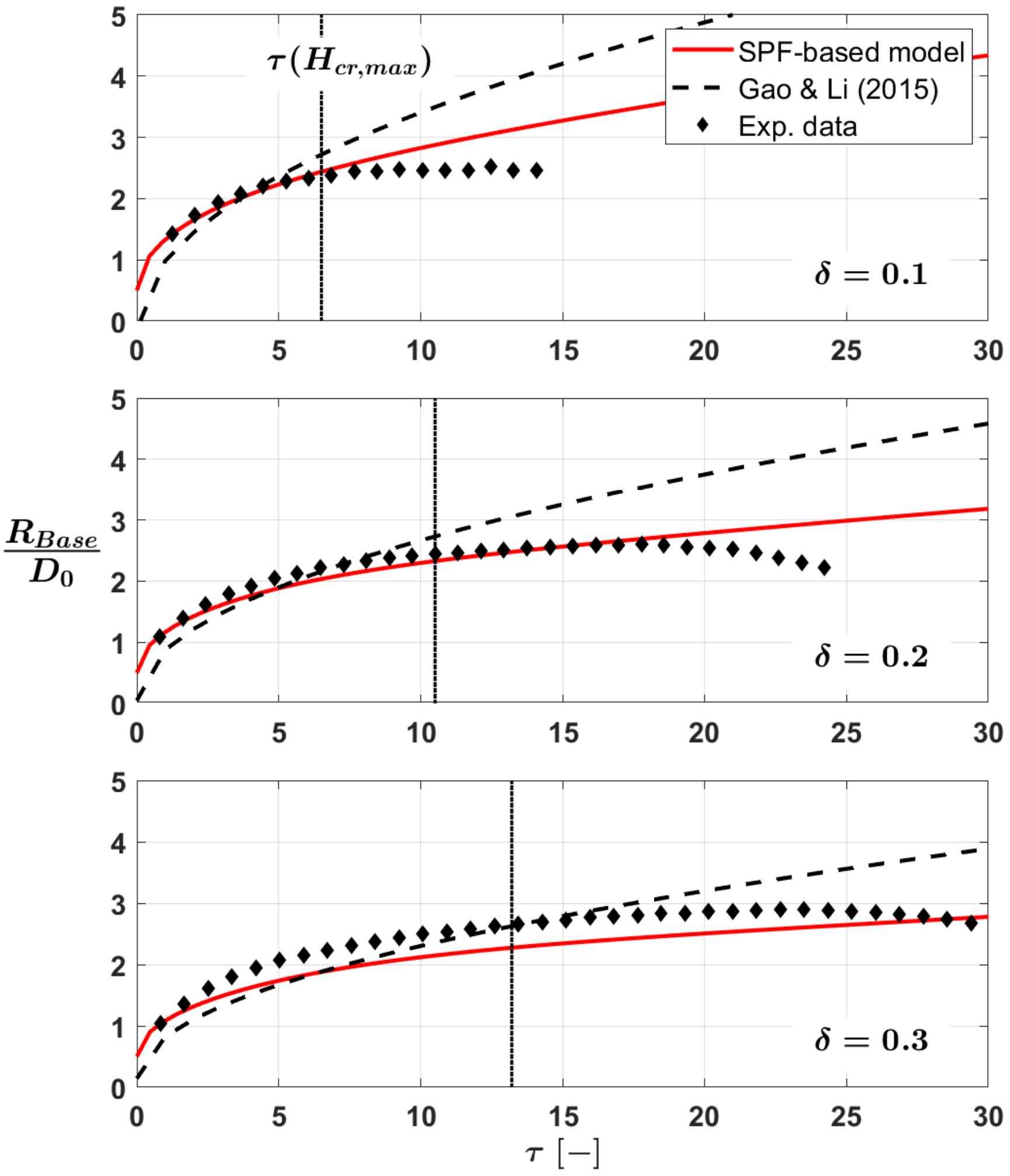}}
			\subcaption{Silicon oil: B10}
   \end{minipage}
	   \begin{minipage}[b]{0.5\linewidth}
			\centering
			{\includegraphics[width=1\linewidth]{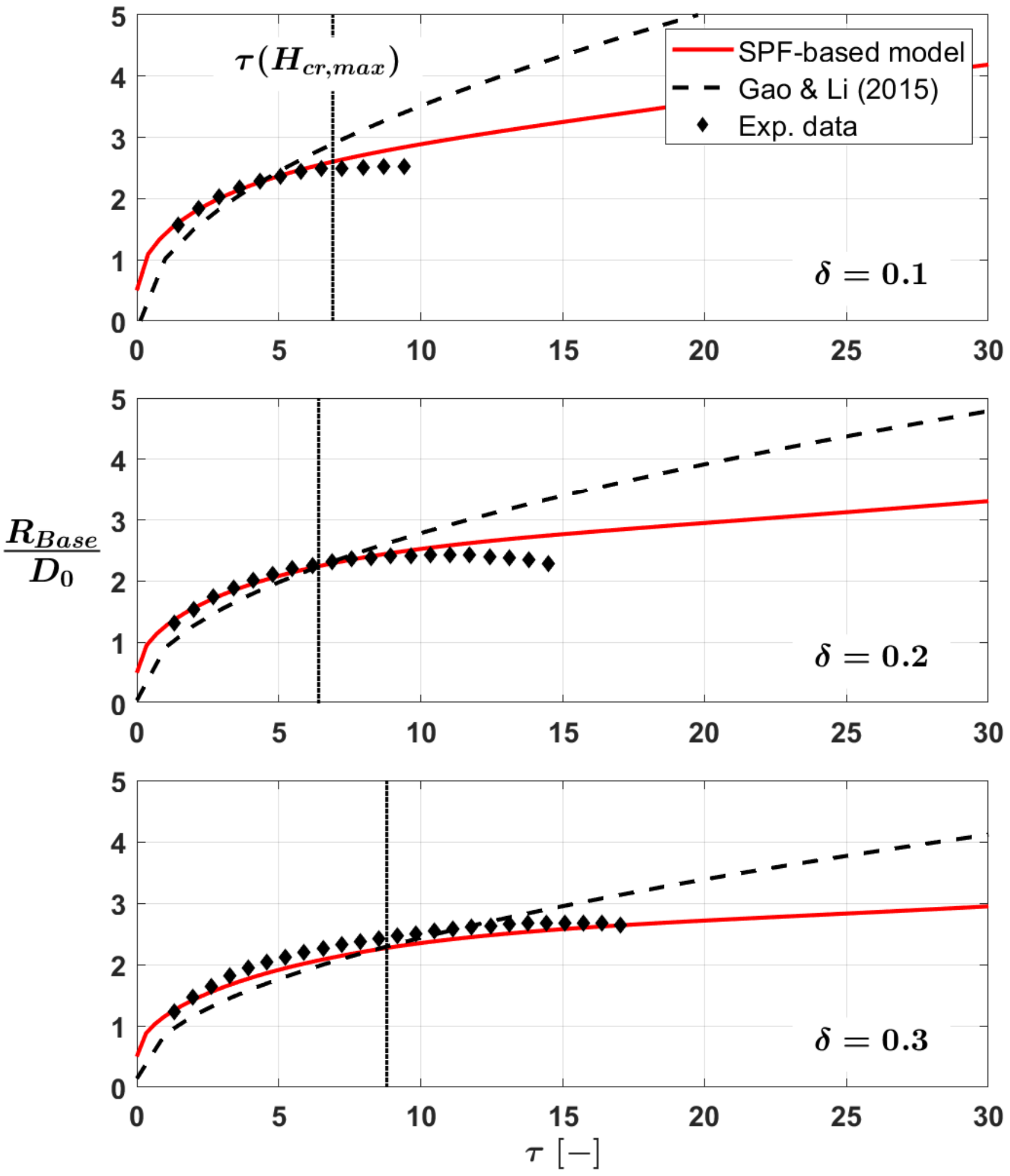}}
			\subcaption{Silicon oil: B3}
   \end{minipage}%
	\begin{minipage}[b]{0.5\linewidth}
			\centering
			{\includegraphics[width=1\linewidth]{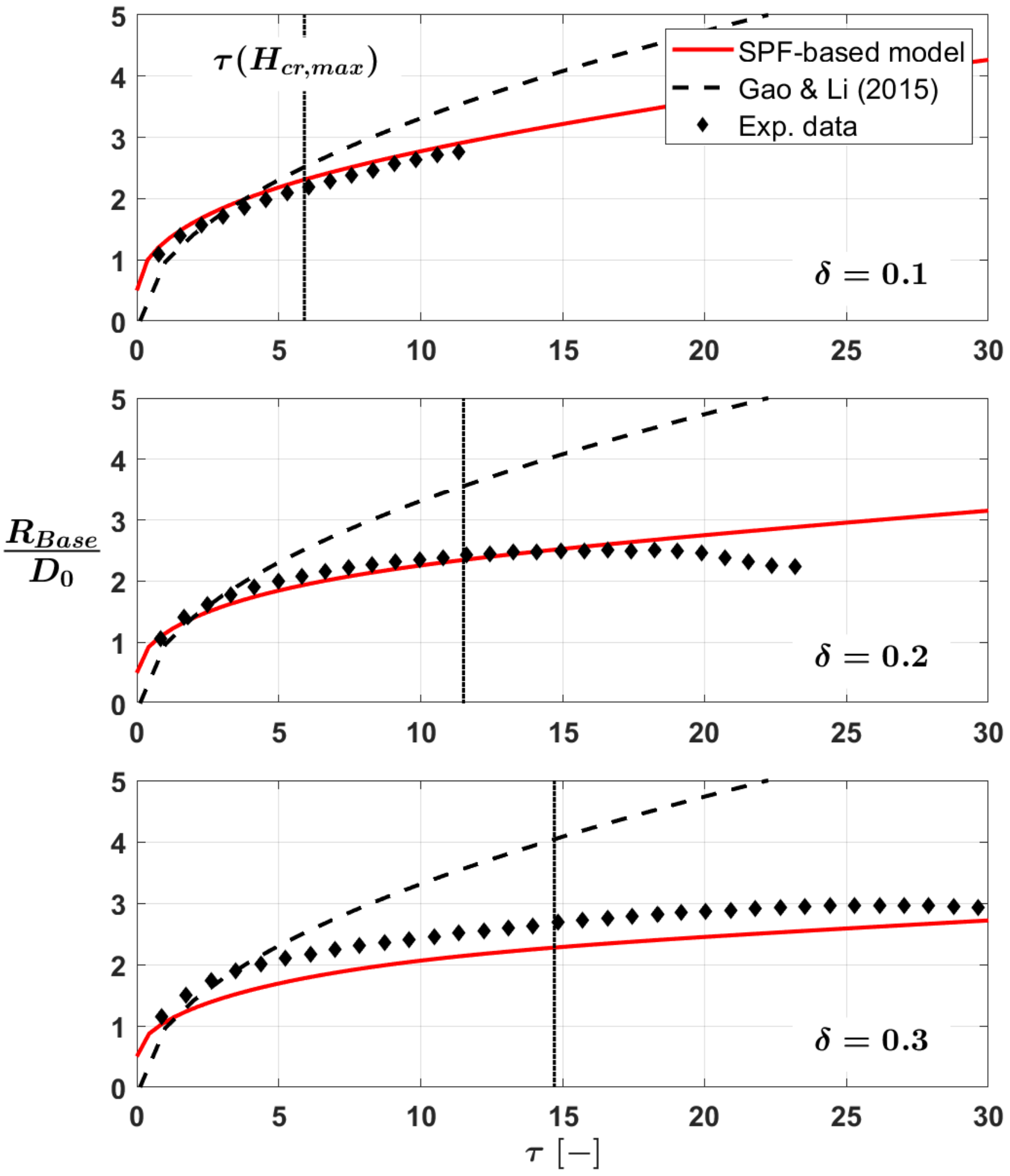}}
			\subcaption{Hyspin}
   \end{minipage}
  \caption{Temporal evolution of the crown base radius $R_{Base}$: comparison between experiments, the Gao and Li model \cite{Gao2015} 
  and the SPF-based approach. The vertical dashed line marks the position of maximum crown's height $H_{cr,max}$, which corresponds to the 
  beginning of the receding phase of the crown. Test conditions are listed in Table \ref{tab:Exp} (see Appendix \ref{app:TC}).}
  \label{fig:Res2}
\vspace{2mm}
\end{figure}

Clearly, the increase in wall-film inertia can also be counterbalanced by increasing the droplet impact velocity $U_0$, as shown in 
Fig.~\ref{fig:inertia}b for the $\delta =0.3$ test case. What is perhaps less intuitive is the decrease in the decay rate of the crown profile-averaged velocity $\bar{u}(t)$ with increasing fluid viscosity. This is due to the fact that the transfer of vertical and horizontal momentum from the impacting droplet to the wall film becomes increasingly less efficient with increasing viscosity. This is clearly visible by comparing the total absolute 
variation in speed from Fig.~\ref{fig:inertia}a to Fig.~\ref{fig:inertia}c: with increasing fluid viscosity the total velocity difference between the 
start and the end of the spreading phase (roughly around $\tau \approx 20-25$ ms) decreases. 
The only noteworthy exception is the B50 case for $\delta = 0.1$, where the initial wall-film thickness ($h_0 = 210 \mu$m) is smaller than the 
displacement thickness ($\delta^* = 277 \mu$m). For this case, potential theory is not applicable and the wall-film decay is controlled by  
the Stokes flow approximation. As shown in Fig.~\ref{fig:inertia}c, the retarding effect due to the presence of the wall is enhanced and results 
in a significant deceleration of the crown's velocity of propagation. As discussed in section \ref{sec:CBB}, this rapid decay of the crown velocity 
is the main factor responsible for the inception of crown bottom breakup (CBB) on thin liquid films.

In order to verify the accuracy of the crown speed predictions, we have compared the temporal evolution of the crown base radius $R_{Base}$ 
with experiments. The results of this exercise are shown in Fig.~\ref{fig:Res2}a to Fig.~\ref{fig:Res2}d for the silicone oils (B50, B10, B3) and 
hyspin. The latter has been added to highlight the reliability of our model in predicting the speed of crown propagation in the event of CBB. For 
each test fluid, the overall agreement of the SPF-model with experimental data is pretty good over the entire duration of the splashing event 
and it provides a significant improvement compared to inviscid models \cite{Yarin1995,Cossali2004,Gao2015}. Two aspects are particularly noteworthy. First, having adopted the Gao and Li's approach as base model, the improvement in the model predictions is obtained without 
introducing any additional empirical correlation. Second, it shows that there is no need to modify the square root time dependence for the 
crown base radius $R_{Base}$ with varying impact conditions, fluid properties and $\delta$, as suggested in \cite{Cossali2004}. The key 
feature for the improved predictions of the stagnation-point flow (SPF) model is that it correctly captures the complex interplay among impact, 
inertial and viscous losses. Finally, Figs.~\ref{fig:Res2}a to \ref{fig:Res2}d corroborate the previous statement on the total duration of a 
splashing experiment, namely it increases with increasing wall-film thickness $\delta$ and fluid kinematic viscosity. Responsible for this 
behaviour is the associated increase in the characteristic time $t_{ref} = D_0 / \bar{u}(t) = D_0 /(\lambda_{AG}(t) U_0)$ for crown propagation. Indeed, as shown in Figs.~\ref{fig:viscosity} and \ref{fig:inertia}, the asymptotic value for the profile-averaged velocity $\bar{u}(t)$ decreases 
with increasing $\delta$ and kinematic viscosity of the fluid.

\subsection{\label{sec:limit}On the limitations of the SPF-model}
The accuracy in the predictions of the SPF-model is hindered by three main factors, namely sliding effects, crown contraction and cavity flow behaviour. This section discusses briefly when these effects become relevant and the reasons for the observed deviations.

\begin{figure}[!b]
       \centering
    \includegraphics[width=0.7\linewidth]{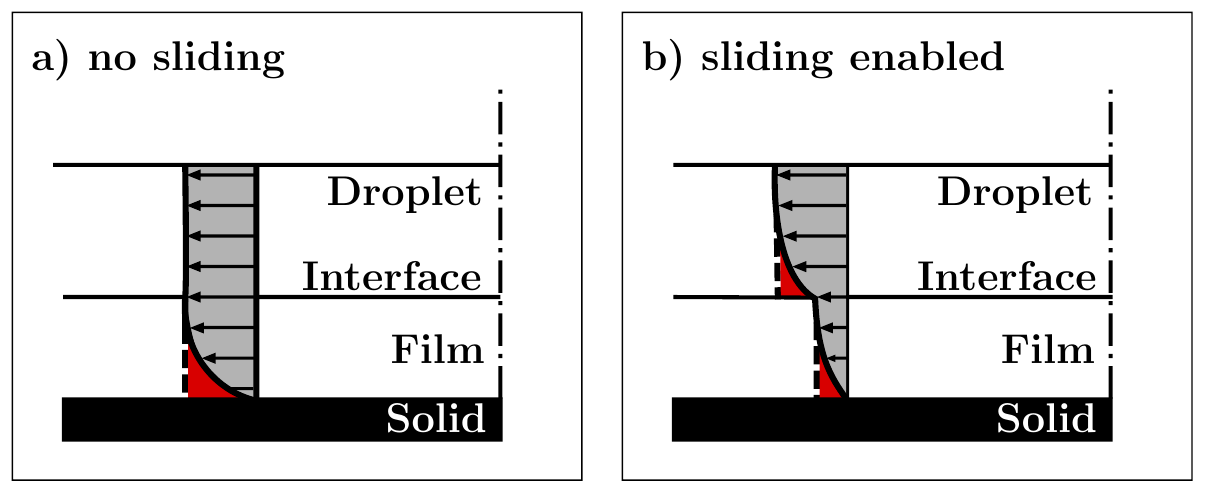}
       \centering
    \caption{Schematic drawing of the velocity profile in the boundary layer for the no-sliding and sliding interface. The grey vector area 
    represents the momentum carried by the lamella, deprived of momentum losses (red areas).}
    \label{fig:BLscheme}
\end{figure}

Sliding effects are typically negligible for low viscous fluids (typically $\nu \leq 5$ mm$^2$/s) or small wall-film thicknesses (typically $\delta 
\leq 0.2$). Outside these limits, the SPF-model is less capable of reproducing exactly the curvature of the experimental $R_{Base}(t)$-profile 
with increasing  $\delta$. Compare, for example, the B3, B10 and B50 curves in Fig.~\ref{fig:Res2}. The SPF-model performs reasonably 
well for B3 up to $\delta = 0.3$ due to its low viscosity. The largest deviations, instead, are observed for the B50, B10 and hyspin cases with 
$\delta = 0.3$ due to their high kinematic viscosities. This is caused by the no-sliding assumption at the interface between the film and droplet 
liquid. The implications of this assumption are schematically illustrated in Fig.~\ref{fig:BLscheme}. The no-sliding assumption implies that the spreading lamella experiences immediately the presence of the solid wall through the no-slip boundary conditions $u_{wall} = 0$. Hence, all momentum losses are confined in the wall film (see Fig.~\ref{fig:BLscheme}a). When sliding is enabled, the lamella can slide on the initially 
resting wall film by introducing an additional boundary condition at the interface, namely $f' = \beta$. Here, $\beta$ represents the 
non-dimensional tangential component of the fluid velocity at the interface and varies from zero (solid boundary) to one (inviscid boundary). 
Subsequently, the velocity gradient is now distributed over a larger distance, which leads to reduced momentum losses from the integration 
of the velocity profile in the boundary layer. This is qualitatively visualised in Fig.~\ref{fig:BLscheme} by the red areas subtended by the 
velocity profiles. Obviously, as discussed in section \ref{sec:Potential} and shown in Fig.~\ref{fig:Mod1}, the presence of the solid wall is 
rapidly felt for low values of $\delta$ due to the fast decay rate of the wall film. Only in these cases, the no-sliding assumption provides a 
good approximation for modelling the spreading of the lamella. 

\begin{figure}[!hb]
  \centering
  \includegraphics[width=0.7\linewidth]{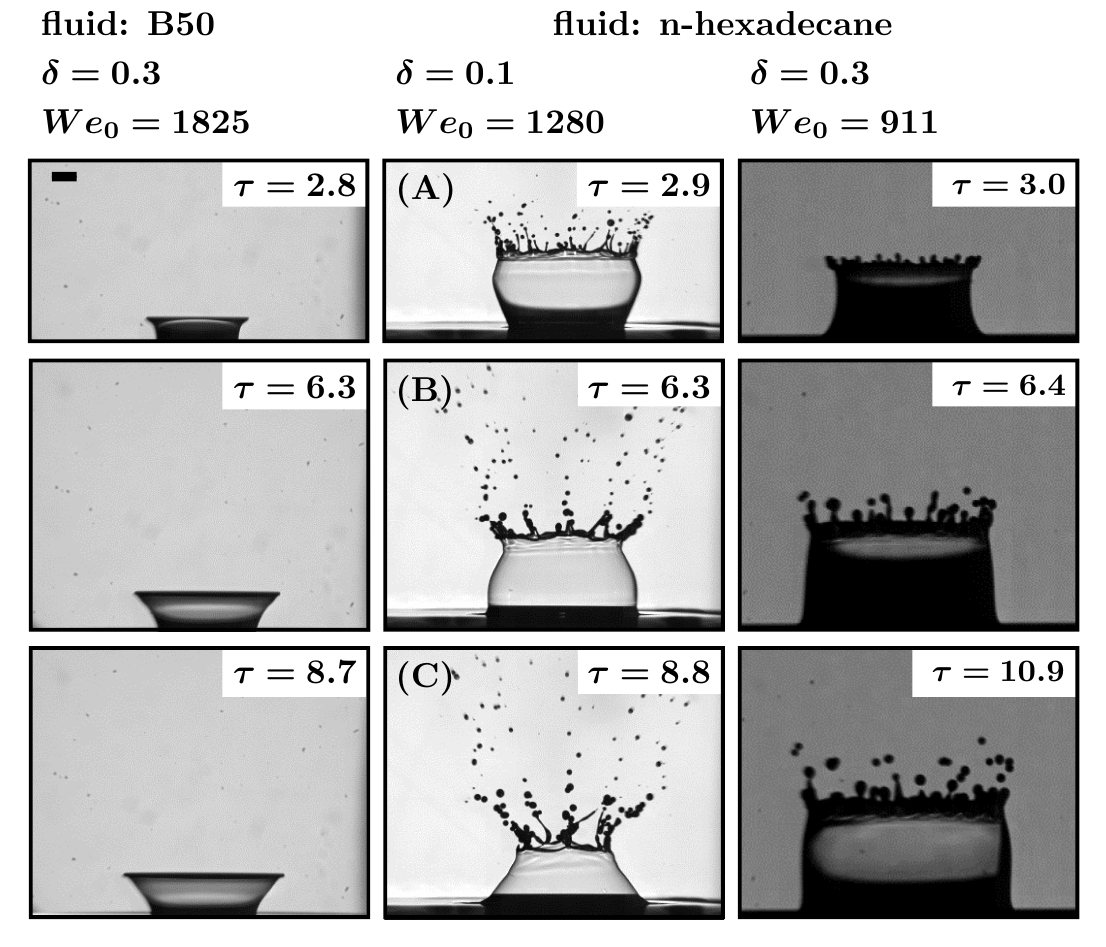}
    \caption{Temporal evolution of the crown morphology in different impact regimes: deposition for B50 and splashing for \textit{n}-hexadecane. 
    Test conditions are listed in Table \ref{tab:Exp} (see Appendix \ref{app:TC}).}
  \label{fig:Res3}
\end{figure}

Wang \cite{Wang1985} demonstrated the existence of a self-similar solution also for a sliding interface by extending the classical Hiemenz 
solution to orthogonal stagnation-point flow against a fluid film, resting against a plane wall. In \cite{Lamanna2019} we adapted the methodology 
of Wang \cite{Wang1985} to a single droplet impinging on a wetted wall. Due to sliding, the integral momentum losses are attenuated in the early phase of spreading. Consequently, a better agreement with experimental data in the range $ 0.3 < \delta < 0.5$ was obtained \cite{Lamanna2019}. The main drawback of Wang's approach is that the non-dimensional sliding velocity $\beta$ must be calculated iteratively at each time step and 
for different initial wall-film thicknesses $\delta$. Hence, unless extremely accurate predictions for the crown radius are required, the classical Hiemenz solution provides a reasonably accurate estimation of the crown evolution without the drawback of an increased computational effort. 
\begin{figure}[!ht]
  \centering
  {\includegraphics[width=0.63\linewidth]{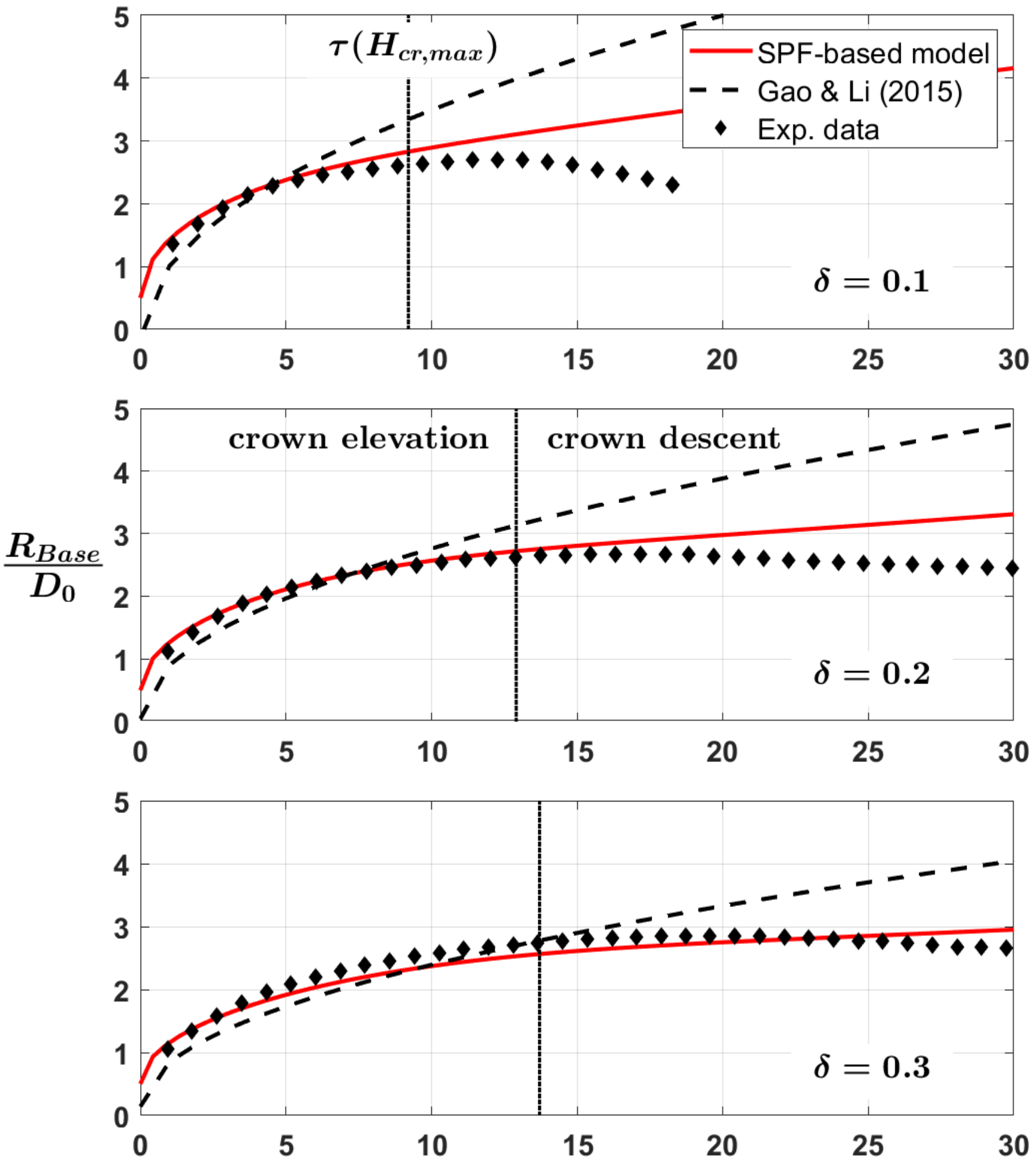}}\\
  \vspace{1mm}
  {\includegraphics[width=0.63\linewidth]{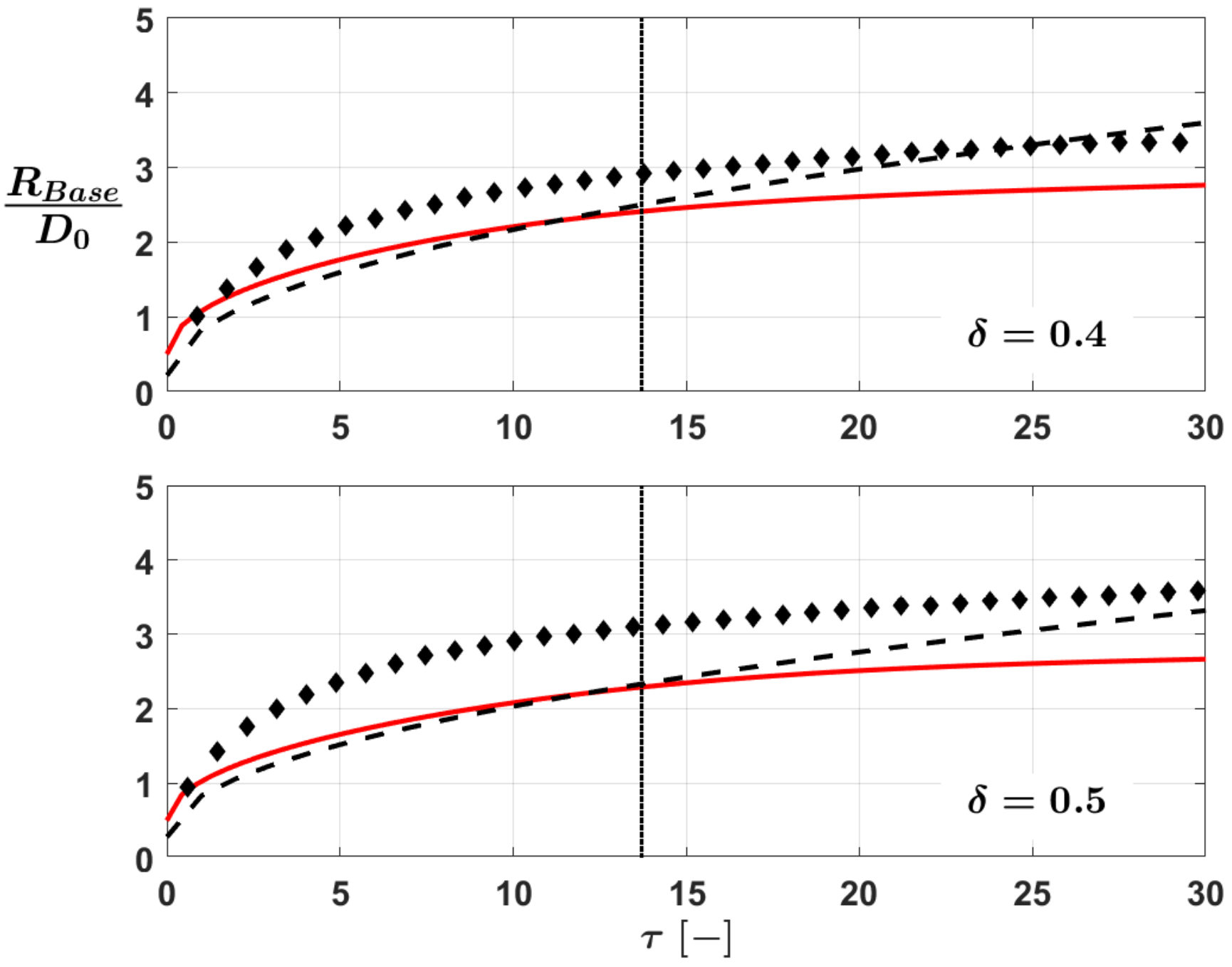}} 
    \caption{Temporal evolution of the crown base radius $R_{Base}$: comparison between experiments, the Gao and Li model \cite{Gao2015} 
  and the SPF-based approach. The vertical dashed line marks the position of maximum crown's height $H_{cr,max}$, which corresponds to 
  the beginning of the receding phase of the crown. Test conditions are listed in Table \ref{tab:Exp} (see Appendix \ref{app:TC}). Fluid: 
  \textit{n}-hexadecane.}
  \label{fig:Res4}
\end{figure}

Strong deviations from the experimental trend may be also observed during the receding phase of a splashing event, in concomitance with 
the occurrence of crown contraction due to capillary forces. Since the effects of surface tension are not encompassed in the stagnation-point 
flow model, it is clear that no accurate prediction of the crown evolution can be expected in this case. Crown contraction has been mainly 
observed in the splashing regime, either for low $We$ numbers (typically $We < 500$) or at high $We$ numbers upon the ejection of secondary droplets. The removal of mass from the crown promotes the restoring action of surface tension and induces changes in crown morphology, as 
shown in Fig.~\ref{fig:Res3} for the \textit{n}-hexadecane test case with $\delta = 0.1$ (second column). As can be seen, crown contraction 
correlates directly with the ejection of secondary droplets and therefore it is temporally delayed with increasing $\delta$, due to the enhanced 
barrier to the onset of splashing with increasing wall-film inertia \cite{Geppert2016,Geppert2017}. Note that crown contraction affects not only 
the rim radius $R_{rim}$, but also its base radius and explains the sudden decrease in the temporal evolution of $R_{Base}$ shown in Fig.~\ref{fig:Res4} for the three test cases with $\delta \leq 0.3$. Note that this trend reproduces systematically also for B3, B10 and hyspin, 
when the droplet impact experiments are performed in the splashing regime. 

In summary, in the splashing regime, the stagnation-point flow model provides an excellent agreement with experiments until the crown 
descent starts. During the receding phase, in fact, capillary effects become important and may lead to the decrease of the crown diameter 
till the emergence of a central jet or a bubble, as discussed in detail in Refs.~\cite{Geppert2016,Geppert2017}. For the experiments considered 
here (e.g.~see Fig.~\ref{fig:Res3} - first column B50), where no ejection of secondary droplets is observed, crown contraction did not occur. 
As a result, accurate predictions of the stagnation-point flow model are obtained even in the receding phase of the crown evolution, as shown 
in Fig.~\ref{fig:Res2}a.


Finally for $\delta \geq 0.5$, the stagnation-point flow model is no longer capable to provide an accurate description of the crown base radius, 
as shown in Fig.~\ref{fig:Res4}. Due to the formation of a cavity, the liquid motion in the impact region is no longer primarily parallel to the wall, 
as assumed in the modelling of the lamella spreading phase in our approach. This statement is corroborated by Lattice-Boltzmann simulations 
in \cite{Mukherjee2007}, where it was shown that, on thicker films, a significant component of the velocity vector is directed downwards to the 
wall. In addition in the case of a cavity, the evolution of the crater diameter is mainly influenced by surface tension and gravity. An accurate 
theoretical model for describing the temporal evolution of liquid cavities can be found in \cite{Roisman2008}.

\subsection{\label{sec:CBB}On the inception of crown bottom breakup}

\begin{figure}
\vspace{2mm}
    \centering
		\begin{minipage}[b]{0.5\linewidth}
		  \centering
 			{\includegraphics[width=1\linewidth]{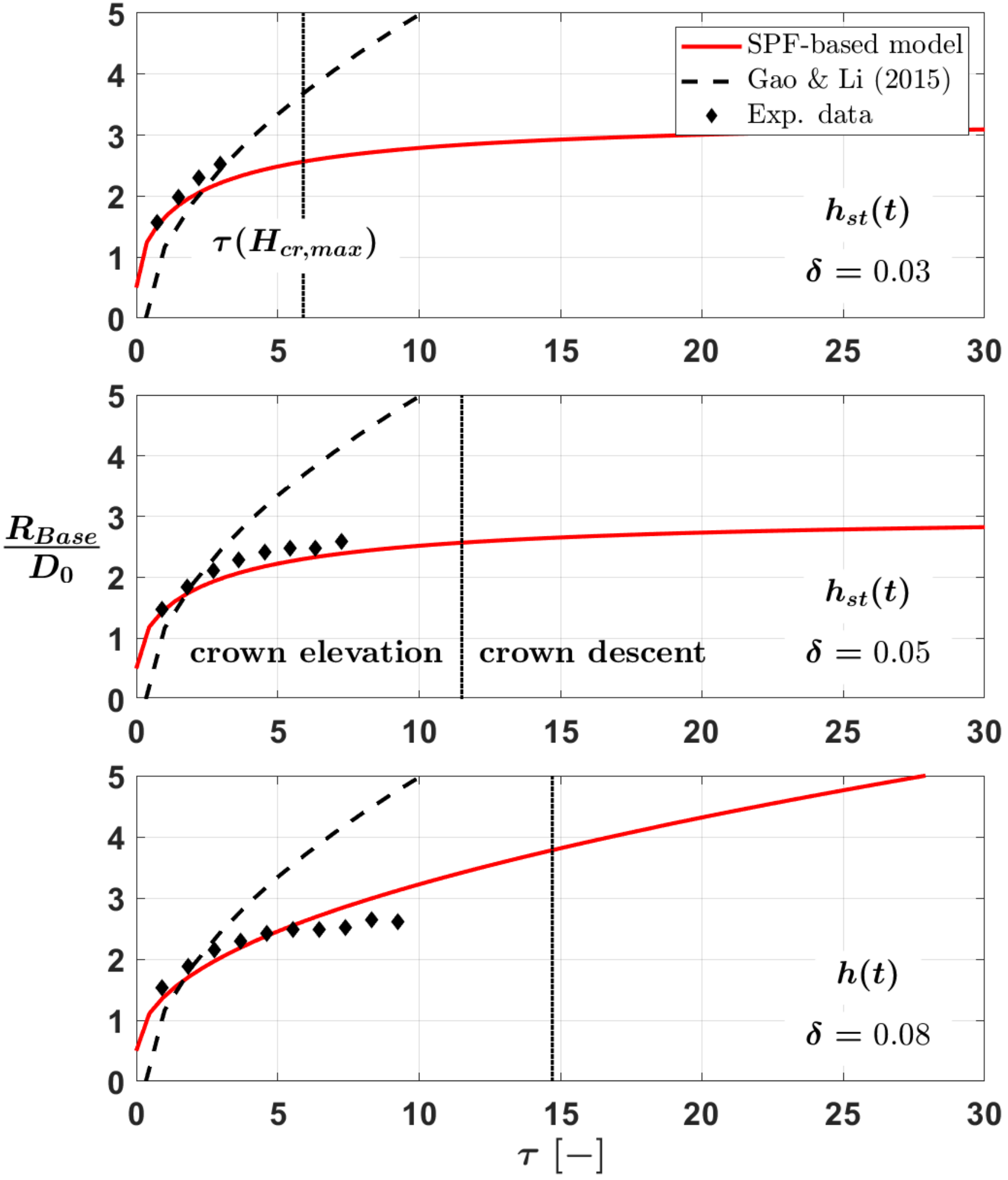}}
		  \subcaption{hyspin}
   \end{minipage}
   \begin{minipage}[b]{0.5\linewidth}
			\centering
			{\includegraphics[width=1\linewidth]{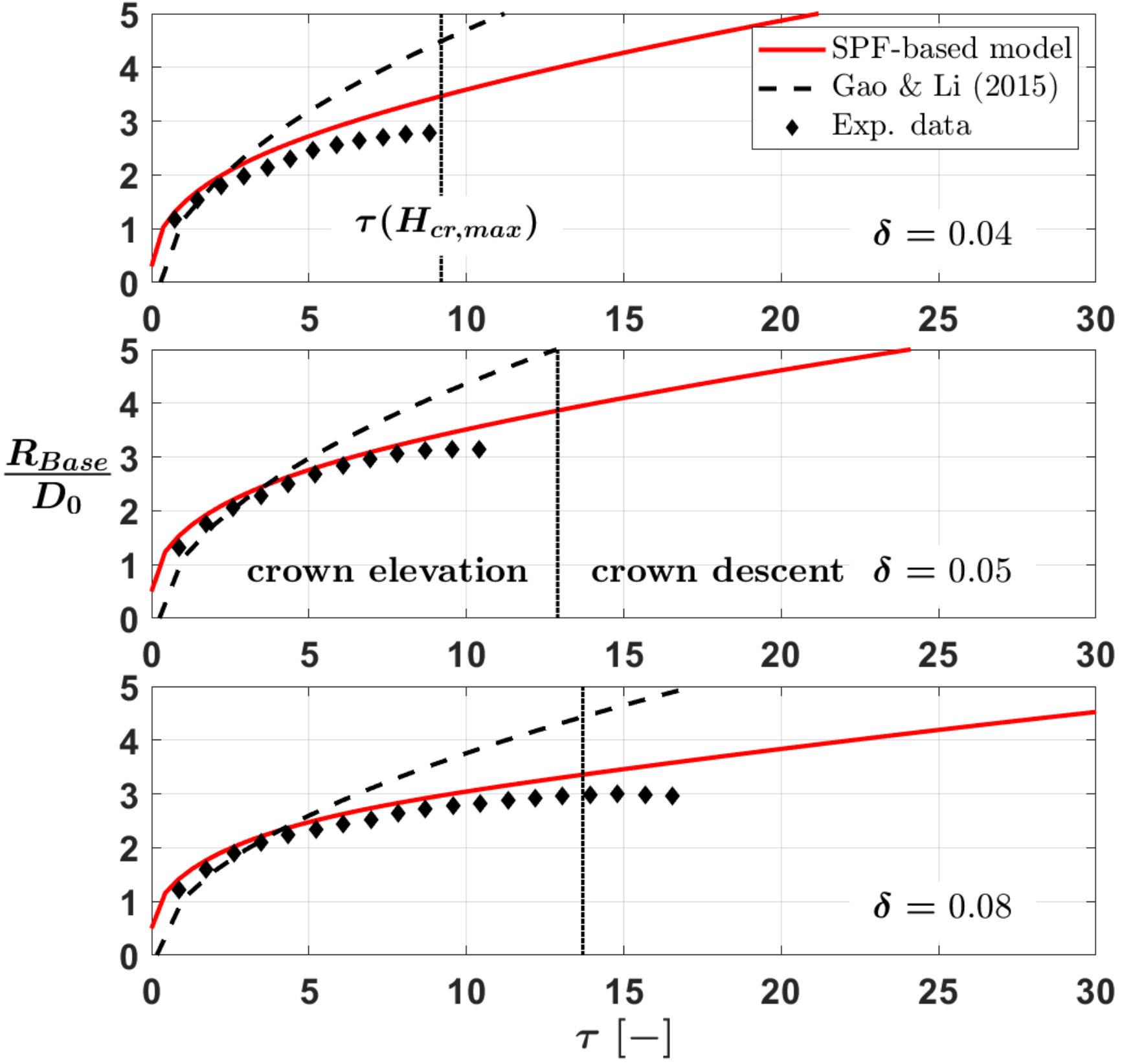}}
			\subcaption{\textit{n}-Hexadecane}
   \end{minipage}
  \caption{Temporal evolution of the crown base radius $R_{Base}$ for $\delta < 0.1$: comparison between experiments, the Gao and Li model 
   \cite{Gao2015} and the SPF-based approach. The vertical dashed line marks the position of maximum crown's height $H_{cr,max}$, which        
   corresponds to the beginning of the receding phase of the crown. Test conditions are listed in Table \ref{tab:Exp} (see Appendix \ref{app:TC}).}
  \label{fig:Res6}
\vspace{2mm}
\end{figure}

\begin{figure}[t]
  \centering
  \includegraphics[width=0.6\linewidth]{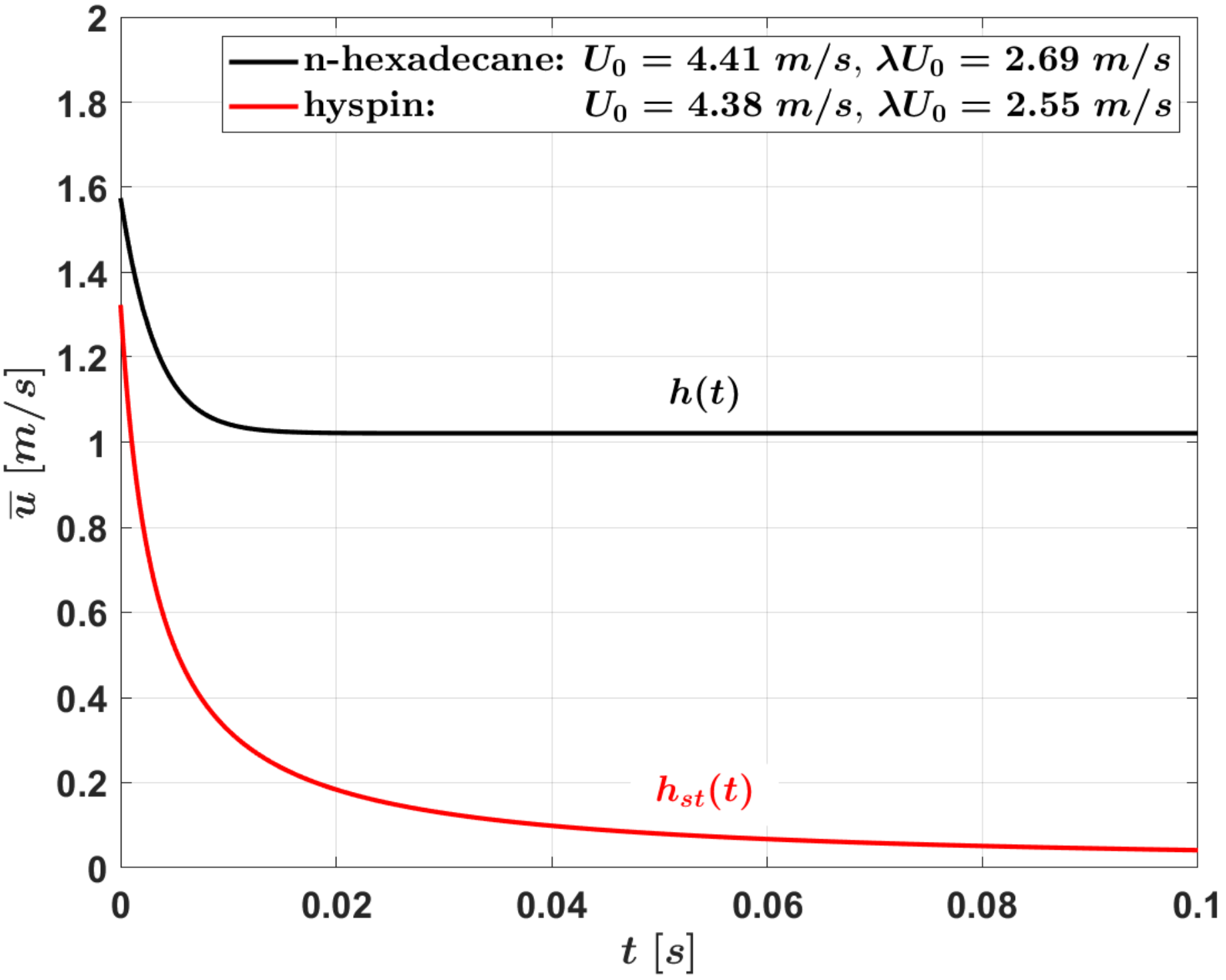}
    \caption{Temporal evolution of the crown profile-averaged velocities for two different test fluids and constant initial film thickness $\delta = 
    0.05$. Note that the Stokes flow solution was employed for hyspin, being $h_0 < \delta^*$. All experimental parameters can be found 
    in Table \ref{tab:Exp} (see Appendix \ref{app:TC}).}	
  \label{fig:Res7}
\end{figure}

As mentioned in section \ref{sec:Intro}, crown bottom breakup (CBB) has been observed experimentally only for very thin wall films ($\delta \leq 0.08$), albeit not consistently for all test fluids. The objective of this section is to investigate whether our analytical approach can provide a sound basis for explaining the occurrence of CBB. As a first step, we verify that the speed of crown propagation is correctly predicted also in the case 
of very thin films. Figure \ref{fig:Res6} shows the temporal evolution of the crown base radius $R_{Base}$ for hyspin and \textit{n}-hexadecane, respectively. As can be seen, the trends discussed in the previous sections are reproduced with satisfactory accuracy for all test cases. For 
$\delta \leq 0.04$, larger discrepancies are observed for \textit{n}-hexadecane due to the premature inception of crown contraction. 
Here it is important to point out that the experimental results confirm the theoretical predictions of section \ref{sec:Viscosity} 
(see Fig.~\ref{fig:inertia}c). For high viscosity fluids, if the initial wall-film height is lower than the displacement thickness, the retarding effect 
of the wall dominates the dynamics of the wall film, leading to a rapid decrease of the crown speed. The theoretical predictions for the 
velocity decay are shown in Fig.~\ref{fig:Res7} for two representative test cases, corresponding to similar impact conditions but different 
fluid viscosities. As can be seen, the Stokes approximation leads to a more pronounced decay of crown speed for hyspin, as confirmed in 
Fig.~\ref{fig:Res6} by comparing the slopes of the associated $R_{Base}$ temporal profiles. The visual inspection of Fig.~\ref{fig:Res6} 
 shows also very clearly the different time scales, at which inertial and viscous forces become predominant. Inertial forces play a major role 
 on the short time scale (up to $\tau \approx 3$) during the process of setting in motion the quiescent wall film. Indeed, all experimental data 
 follow very closely the Gao and Li curve \cite{Gao2015}. Instead, viscous losses during the crown propagation need a longer time scale to 
 manifest themselves. This effect reproduces very consistently in our database with a systematic deviation from the Gao and Li curve on 
 the longer time scale. Note that the deviation point in time varies with film thickness. For $\delta < 0.1$, it occurs around $\tau \approx 3-4$ 
 (see Fig.~\ref{fig:Res6}); while for $\delta \geq 0.1$ it occurs between $5 < \tau < 10$ depending on fluid viscosity and film thickness, as 
 shown in Fig.~\ref{fig:Res2}. These experimental findings corroborate our theoretical predictions that viscous losses become increasingly predominant with decreasing film thickness.
 
\begin{figure}[h]
  \centering
  \includegraphics[width=0.6\linewidth]{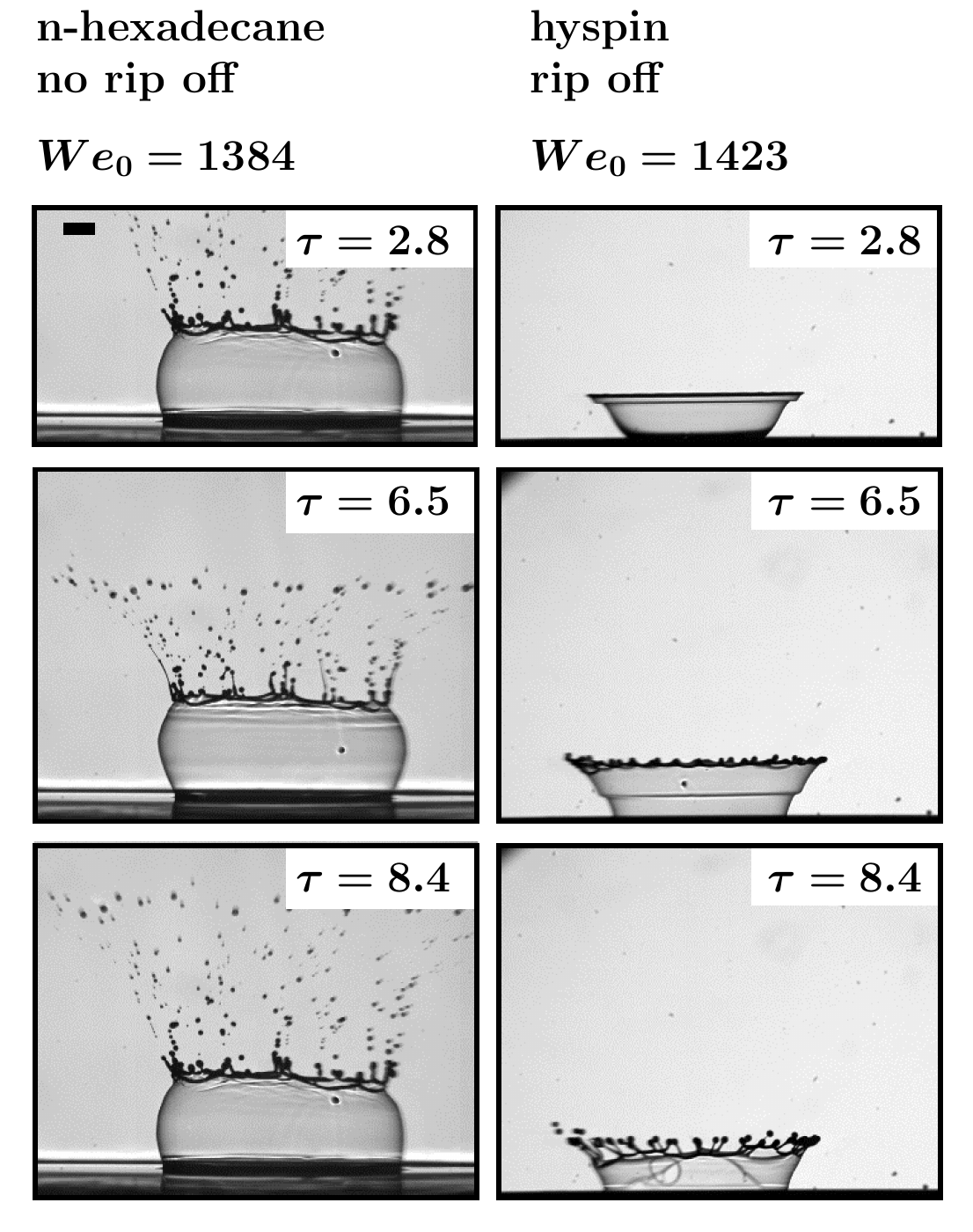}
    \caption{Temporal evolution of the crown morphology for impact on thin wall film: $\delta = 0.05$. Fluids: hyspin and \textit{n}-hexadecane. 
    Test conditions are listed in Table \ref{tab:Exp} (see Appendix \ref{app:TC}).}	
  \label{fig:Res8}
\end{figure}

Finally, it is important to point out that viscous losses alone cannot explain the onset of crown bottom breakup (CBB). As pointed out by 
Rozhkov et al.~\cite{Rozhkov2004,Rozhkov2010}, the evolution of the crown morphology plays also a key role for explaining the inception 
of crown bottom breakup (CBB). Figure \ref{fig:Res8} illustrates the evolution of the crown morphology for droplet impact on a thin wall film 
($\delta = 0.05$) for the same test conditions of Fig.~\ref{fig:Res7}. While the \textit{n}-hexadecane crown experiences a contraction, the 
hyspin crown is streching during the entire duration of the splashing event. In agreement with the theoretical predictions of Rozhkov et al.~\cite{Rozhkov2004,Rozhkov2010}, CBB is observed systematically only for impacts with hyspin. Based on these experimental 
observations, in this work we extended the model of Rozhkov et al.~\cite{Rozhkov2004,Rozhkov2010}, initially developed for drop impact 
on a small solid target, to droplet impact on wetted walls of finite extent. The starting point is to assimilate the upward expansion of the 
crown lamella to the ejection of a liquid sheet from a point source with velocity $v_{s}$ and thickness $h_{lam}$. Further we assume that, 
at the crown base, the thickness of the lamella $h_{lam}$ and its ejection velocity $v_s$ are approximately equal to the wall-film height 
[i.e.~$h_{lam}(t) \approx h(t)$] and the average speed of propagation of the crown base radius [i.e.~$v_s(t) \approx \bar{u}(t)$]. The local 
mass flow rate $q(t)$ at the position $R_{Base}$, entering the ejected crown lamella, is defined as equal to the amount of liquid that flows 
through a circular contour of radius $R_{Base}$ per unit of time. The dimensionless mass flow rate can be then expressed as
\begin{equation}
Q_s = \frac{2 \pi R_{Base} \bar{u}(t) h(t)}{ \pi D_0^2 U_0/6}
\label{eq:res1}
\end{equation}
where the expression ($\pi \rho D_0^2 U_0 / 6$) represents a reference mass flow rate at the impact, being equal to the ratio of drop mass 
($\pi \rho D_0^3 / 6$) to the experimental initial time scale $D_0/U_0$. Obviously, since the flow can be considered incompressible, the 
density has been simplified in Eq.~(\ref{eq:res1}). Note that, if both momentum losses and the decay in wall-film thickness are neglected, 
this will automatically lead to a significant overestimation of the mass flow rate entering the ejecta sheet.
  
  \begin{figure}[t]
  \centering
  \includegraphics[width=0.4\linewidth]{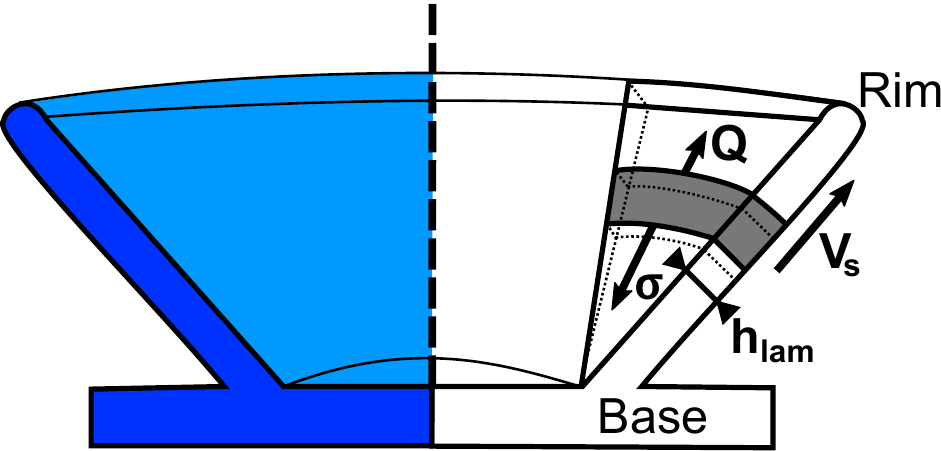}
    \caption{Schematic drawing of the crown lamella, stretching upwards during the elevation phase. The grey control volume shows the 
    counteracting action between inertial forces and surface tension.}	
  \label{fig:Res8.5}
\end{figure}

\begin{figure}[t]
 \centering
  \includegraphics[width=0.67\linewidth]{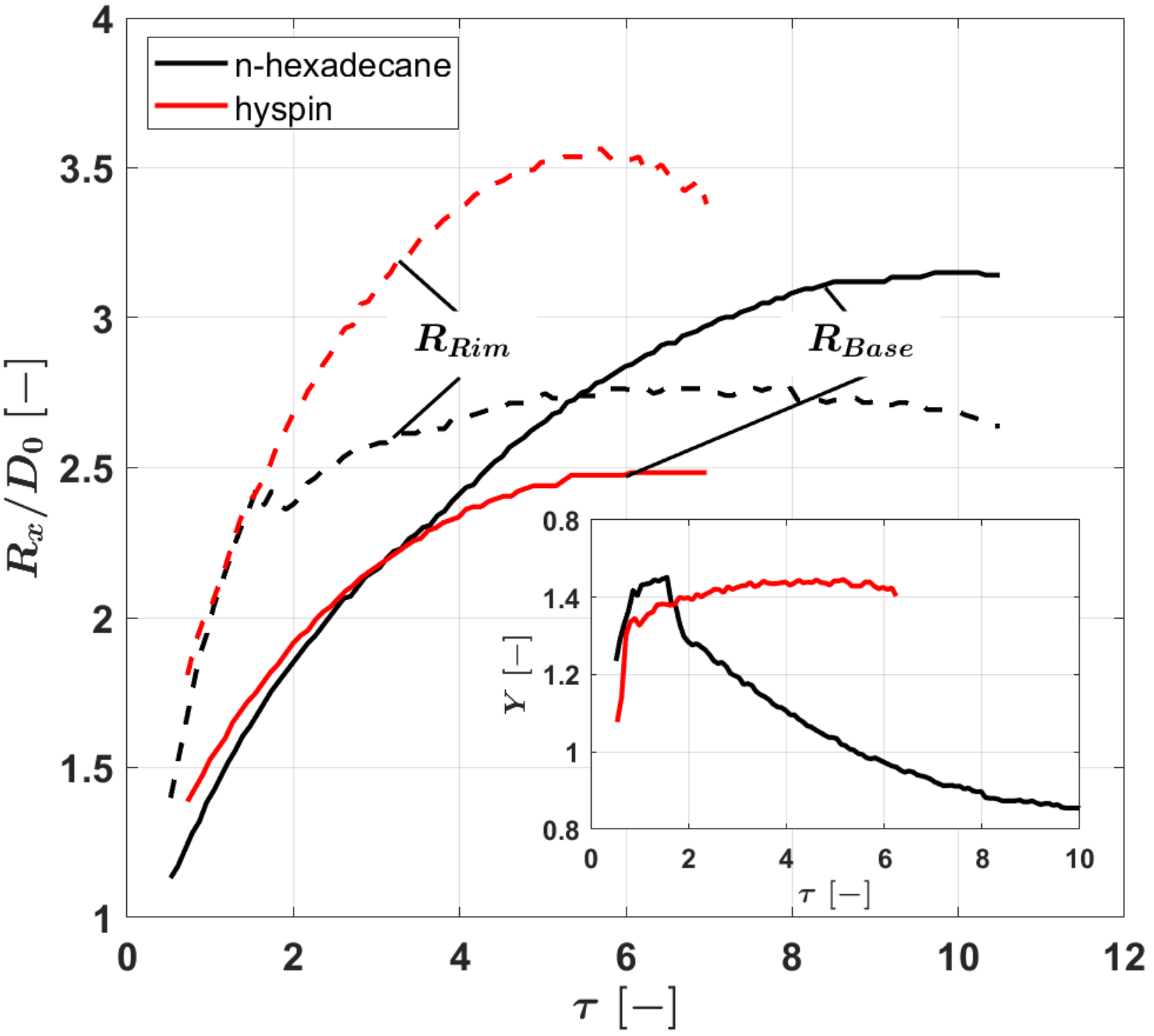}
  \caption{Temporal evolution of the crown rim ($R_{Rim}$) and base radius ($R_{Base}$) for two different test fluids and constant initial film 
  thickness $\delta = 0.05$. The insert depicts the associated temporal variation of the spreading factor $Y$. All other experimental parameters 
  can be found in Table \ref{tab:Exp} (see Appendix \ref{app:TC}).}	
  \label{fig:Res9}
\end{figure}

Typically during a splashing event, any liquid element entering the crown lamella experiences a thinning in time due to its radial and azimuthal 
extension, as schematically shown in Fig.~\ref{fig:Res8.5}. Due to mass conservation, the local flow rate must necessarily decrease. For a 
stretching lamella, Rozhkov et al.~\cite{Rozhkov2004,Rozhkov2010} postulated a universal function for the decrease of the dimensionless 
mass flow rate, according to
\begin{equation}
Q(\tau) = \frac{Q_s}{1 + Y \frac{d}{d \tau} \left (\frac{1}{V_s(\tau)} \right )}
\label{eq:res2}
\end{equation}
where $V_s(\tau) = v_s(t)/U_0$ is the dimensionless ejection velocity, obtained from our analytical solution. For the test cases plotted in Fig.~\ref{fig:Res7}, the functional dependence for $V_s(\tau)$ and its temporal derivative is reported in Appendix \ref{App:CBB}. The parameter 
$Y = R_{Rim}/R_{Base}$ represents the lamella spreading factor and is obtained experimentally. The temporal variation of the crown rim and 
base radius together with the associated values of the spreading factor $Y$ are shown in Fig.~\ref{fig:Res9} for the same two representative 
test cases ($\delta = 0.05$) of Fig.~\ref{fig:Res7}. When $Y$ decreases, fluid elements in the \textit{n}-hexadecane crown 
experience a narrowing of their radial cross section during a significant portion of the crown evolution, and hence an increase in local mass 
flow rate. The opposite trend is observed for fluid elements in the hyspin crown. The progressive decrease of ejection velocity and local mass 
flow rate in the hyspin crown leads to the creation of metastability zones within the stretching lamella due to the increasing unbalance between 
inertia and surface tension forces. In non-dimensional terms, this condition is met when the local $We$ number ($We_{local} = \rho h_s v_s^2 / \sigma$) in the expanding lamella drops below one. Taking into account the continuity equation, $We_{local}$ can be expressed as follows \cite{Rozhkov2004,Rozhkov2010} 
\begin{equation}
We_{local} = \frac{ \left (\frac{1}{24} We_0 \right )V_s(\tau) Q(\tau)}{Y(\tau)}.
\label{eq:res3}
\end{equation}

\begin{figure}[t]
  \centering
  \includegraphics[width=0.67\linewidth]{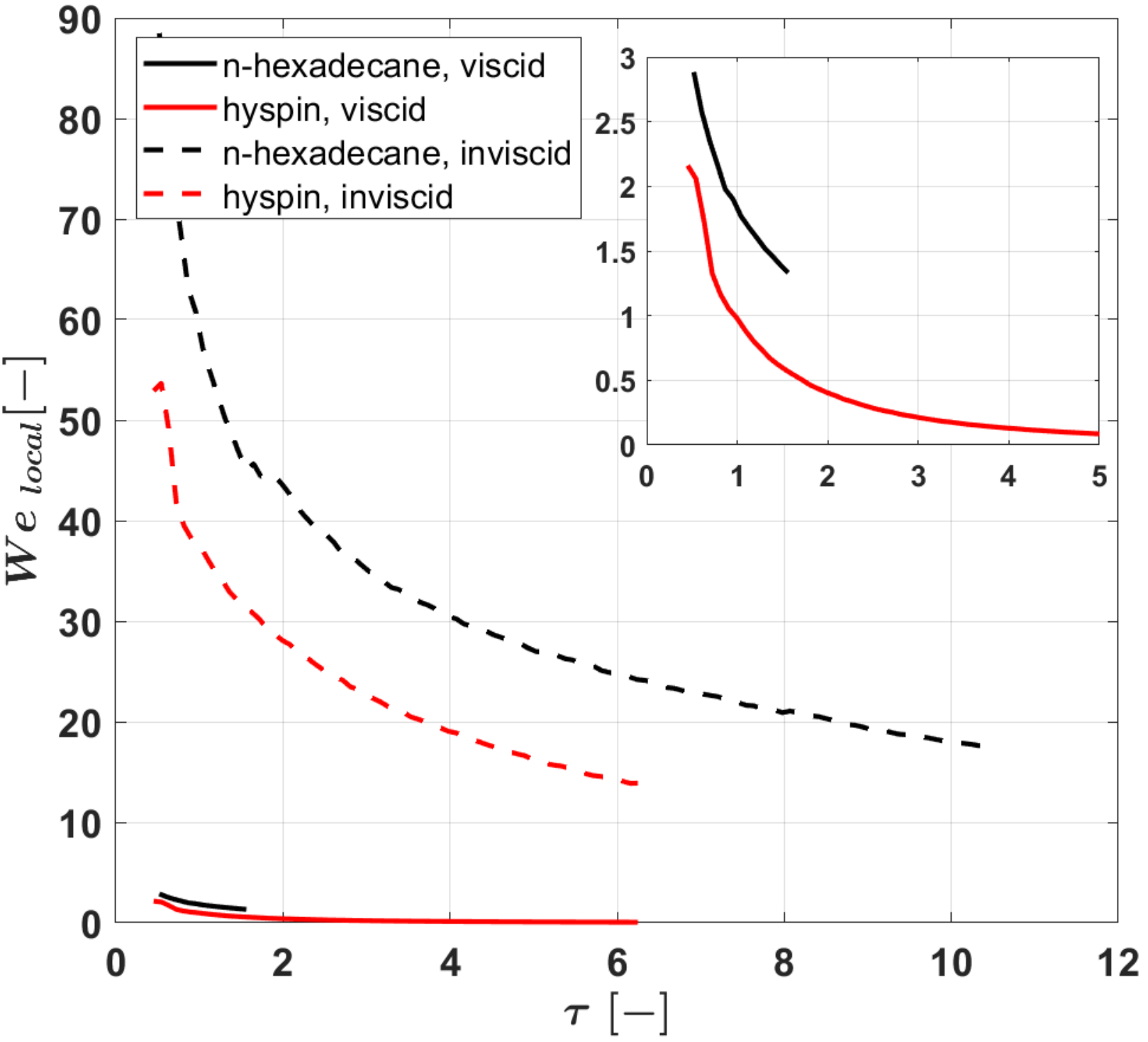}
    \caption{Temporal evolution of the local Weber number $We_{local}$ for two different test fluids and constant initial film thickness 
    $\delta = 0.05$, as predicted by the SPF-model and the inviscid solution with $\bar{u}(t)= \lambda U_0 = const$. The insert highlights the     
    associated decay of $We_{local}$ for the SPF-model. All other experimental parameters can be found in Table \ref{tab:Exp} (see Appendix \ref{app:TC}).}	
  \label{fig:Res10}
\end{figure}

The temporal evolution of $We_{local}$ is shown in Fig.~\ref{fig:Res10} for the same representative test cases of Fig.~\ref{fig:Res9}. 
In the metastability zone (i.e.~$We_{local} << 1$), any small disturbance cannot be transported away by the flow, thus yielding to a rapid destabilisation (rupture) of the lamella. Indeed, there is a direct correspondence between the experimental images in Fig.~\ref{fig:Res8} 
and the temporal evolution of the local We number (or equivalently local dimensionless flow rate $Q$), namely the occurrence of CBB 
corresponds to the condition of vanishing local flow rate $Q \rightarrow 0$ (or $We_{local} \rightarrow 0$). By correlating the temporal 
evolution of the spreading factor $Y$ with $We_{local}$, it is also immediately clear why no CBB is observed for \textit{n}-hexadecane. 
During the stretching phase of the \textit{n}-hexadecane crown (till approximately $\tau \approx 1.5$ in Fig.~\ref{fig:Res9}), $We_{local}$ 
remains always above one (see insert in Fig.~\ref{fig:Res10}), corresponding to a dynamically stable evolution of the crown. Note that the 
exact instant in time for the onset of CBB is difficult to predict theoretically, since it is connected to local random disturbances (e.g.~dust, 
impurities, secondary droplets impinging on the lamella) that trigger the propagation of the rupture at the Taylor-Culick velocity \cite{Taylor1959,Pandit1990}, which is considerably larger than the local flow velocity in the lamella $v_s(t)$.  Note that the values of 
$We_{local}$ for the \textit{n}-hexadecane crown have been plotted only for the stretching phase of the lamella, since no model is available 
to describe the local increase in mass flow rate for a contracting crown. Finally, Fig.~\ref{fig:Res10} shows also the temporal evolution of 
$We_{local}$, as predicted by inviscid models. The curve is obtained by using Eqs.~(\ref{eq:res1})-(\ref{eq:res3}) and setting $\bar{u}(t)= 
\lambda U_0$. As can be seen, inviscid model that neglect viscous losses during the propagation of the crown (kinematic discontinuity) 
cannot reproduce the creation of metastability areas within a stretching lamella, thus leaving the occurrence of CBB basically unexplained.
 
\section{Conclusions}
This paper discusses a new approach for modelling the crown propagation, based on analytical solutions of the Navier-Stokes equations 
for stagnation-point flow (SPF-model). As a starting point, droplet impact on a wet substrate is ideally divided into two sub-processes, namely 
impulse transfer (phase a) and lamella spreading (phase b). As a result of the collision, momentum is transferred from the impinging droplet 
to the wall film (phase a), thereby causing a decrease in wall-film thickness with time $h(t)$. The decay in wall-film height is modelled by 
assuming that the flow inside the droplet resembles the potential flow of a stagnation-point problem. Right after the impact, we assume that 
the liquid-droplet and the wall-film fluids merge perfectly and start spreading radially outwards. The profile-averaged velocity distribution 
$\bar{u}_x(t)$ increases linearly along the $x$-axis from zero at the stagnation-point ($x =0$) to its maximum value $\bar{u}(t)$ 
at the crown (kinematic discontinuity). The latter decreases progressively in time due to viscous losses and is estimated from the Hiemenz's 
boundary layer solution for a plane stagnation-point flow. Finally, $\bar{u}(t)$ is inserted in the inviscid model for crown propagation of Gao and 
Li \cite{Gao2015}. Contrary to all previous theories \cite{Yarin1995,Cossali2004,Gao2015}, this modification includes, \emph{de facto}, the 
deceleration due to momentum losses in the modelling of crown propagation.

Our analysis shows that, during the spreading phase (phase b), viscous losses are negligible only in the early phase of crown propagation 
and become increasingly important with reducing film height. This enables a smooth transition from the inertia-driven to the viscous-controlled 
regime of crown propagation. Indeed, the overall agreement of the stagnation-point flow model with experimental data is pretty accurate over the entire duration of the splashing event and it provides a significant improvement compared to inviscid models. Overall, viscous 
losses affect the temporal evolution of the crown base radius in two ways: directly by causing momentum losses during the spreading of the 
lamella and indirectly in the impulse transfer from the droplet to the wall film. For very viscous fluids, the transfer of vertical and horizontal 
momentum from the impacting droplet to the wall film is inhibited due to the high impact losses. The latter can also be incremented by increasing 
the wall-film inertia, i.e.~increasing $\delta$. This gives rises to a complex interplay between impact an viscous losses: the higher the impact 
losses, the lower the total viscous losses in the spreading phase and vice versa. This counterbalancing of impact and viscous losses explains 
why the crown base radius evolution was found to be independent of $\delta$ for a large part of the crown dynamics.
 
Finally, our analytical solution for the crown's speed of propagation and wall-film thickness also enables to calculate the time dependent variation 
of local mass flow rate and local Weber number $We_{local}$ in the ejecta sheet. Here, it is important to point out that, if the initial wall-film thickness $h_0$ is smaller than the displacement thickness $\delta^*$, the Stokes flow approximation must be employed for modelling the wall-film decay, because potential theory is no longer applicable to model the impulse transfer from the impinging droplet. It is found that, for very 
thin wall films, zones of metastability can be formed in the crown lamella, corresponding to the condition $We_{local} << 1$. In these metastable zones, any small disturbance cannot be transported away by the flow, thus causing the rupture of the lamella and the creation of a web-like 
structure. The velocity of propagation of the rupture is the Taylor-Culick velocity, which is significantly larger than the local flow velocity in the 
lamella $v_s(t)=\bar{u}(t)$, and leads to the rapid disintegration of the lamella close to the crown base.

The accuracy in the predictions of the SPF-model is hindered by three main factors, namely sliding effects, crown contraction and cavity flow behaviour for $\delta \geq 0.5$. Despite these limitations, the proposed modelling strategy provides a significant step forward in the prediction 
and understanding of crown propagation on wetted walls. First, it provides a straightforward explanation for the different timescales observed numerically on the spreading rate of the corolla \cite{Marcotte2019}. Second, it is capable to correctly reproduce the complex interplay among 
impact, inertial and viscous forces in controlling the evolution of the crown base radius. Third, it paves the way for understanding and predicting 
the occurrence of crown bottom breakup. Fourth, the SPF-methodology can be easily extended to analyse the effect of sliding for droplet 
impinging on wall films of different viscosity. This is a necessary step to assess how the viscosity ratio affects the overall crown spreading and splashing dynamics. \\

\textbf{Acknowledgments}:
The authors wish to thank the Deutsche Forschungsgemeinschaft (DFG) for financial support in the framework of the projects 
LA 2512/2-1, WE 2549/24-1 and GRK 2160/1 ``Droplet Interaction Technologies'' (DROPIT).

\section*{Nomenclature}
\begin{tabbing}
\hspace*{2.5 cm}  \= \hspace*{8 cm}  \= \hspace*{3 cm} \kill
{\rm $a$} \> {\rm momentum per unit length, $\lambda U_0/\pi D_0$} \> {[$1$/$s$]}\\
{\rm $C_1$} \> {\rm non-dim. variable, $(2/3\delta)^{1/4}$} \> {[-]}\\
{\rm $C$} \> {\rm non-dim. variable, $(2\lambda^2/3\delta)^{1/4}$} \> {[-]}\\
{\rm $C(t)$} \> {\rm  time-dependent variable, $(2\lambda_{AG}^2/3\delta)^{1/4}$} \> {[-]}\\
{\rm $D_0$} \> {\rm droplet diameter} \> {[$m$]}\\
{\rm $f$} \> {\rm non-dim. stream function, $\psi/x\sqrt{\nu a}$} \> {[-]}\\
{\rm $f',f'',f'''$} \> {\rm first, second, third derivative of $f$} \> {[-]}\\
{\rm $\bar{f}'$} \> {\rm profile-averaged non-dim. velocity} \> {[-]}\\
{\rm $g$} \> {\rm non-dim. stream function, $g = f(\xi) \sqrt{t}$} \> {[-]}\\
{\rm $g',g'',g'''$} \> {\rm first, second, third derivative of $g$} \> {[-]}\\
{\rm $G$} \> {\rm strength of the Stokes flow} \> {[$1$/$(ms)$]}\\
{\rm $H_{cr,max}$} \> {\rm maximum crown height} \> {[$m$]}\\
{\rm $h_0$} \> {\rm initial wall-film thickness} \> {[$m$]}\\
{\rm $h(t)$} \> {\rm wall-film thickness decay rate} \> {[$m$]}\\
{\rm $h_{st}(t)$} \> {\rm Stokes decay rate} \> {[$m$]}\\
{\rm $h_{lam}(t)$} \> {\rm lamella thickness} \> {[$m$]}\\
{\rm $K$} \> {\rm splashing factor, $We_{0}^{0.5} Re_{0}^{0.25}$} \> {[-]}\\
{\rm $n$} \> {\rm exponent} \> {[-]}\\
{\rm $q(t)$} \> {\rm mass flow rate} \> {[$kg$/$s$]}\\
{\rm $Q_s$} \> {\rm non-dim. mass flow rate} \> {[-]}\\
{\rm $Q(\tau)$} \> {\rm time-dep. non-dim. mass flow rate} \> {[-]}\\
{\rm $R_{Base}$} \> {\rm crown base radius} \> {[$m$]}\\
{\rm $R_{Rim}$} \> {\rm crown radius at upper rim} \> {[$m$]}\\
{\rm $Re_0$} \> {\rm initial Reynolds number, $U_0 D_0$/$\nu$} \> {[-]}\\
{\rm $t$} \> {\rm time} \> {[$s$]}\\
{\rm $t_{ref}$} \> {\rm reference time} \> {[$s$]}\\
{\rm $t_{end}$} \> {\rm duration of one experiment} \> {[$s$]}\\
{\rm $U_0$} \> {\rm initial droplet velocity} \> {[$m$/$s$]}\\
{\rm $\bar{u}(t)$} \> {\rm profile-averaged crown velocity} \> {[$m$/$s$]}\\
{\rm $\bar{u}_x(t)$} \> {\rm profile-averaged velocity distribution} \> {[$m$/$s$]}\\
{\rm $u_{\infty} $} \> {\rm velocity outside boundary layer, $\lambda U_0$} \> {[$m$/$s$]}\\
{\rm $u,v$} \> {\rm velocity components} \> {[$m$/$s$]}\\
{\rm $v_s$} \> {\rm point source velocity} \> {[$m$/$s$]}\\
{\rm $V_s(\tau)$} \> {\rm non-dim. ejection velocity} \> {[-]}\\
{\rm $We_0$} \> {\rm initial Weber number, $\rho U_0^2 D_0$/$\sigma$} \> {[-]}\\
{\rm $We_{local}$} \> {\rm local Weber number} \> {[-]}\\
{\rm $x,y$} \> {\rm coordinates} \> {[-]}\\
{\rm $Y$} \> {\rm lamella spreading factor, $R_{Rim}/R_{Base}$} \> {[-]}\\
\\{\rm Greek letters}\\
\\
{\rm $\beta$} \> {\rm non-dim. tangential velocity} \> {[-]}\\
{\rm $$} \> {\rm component at interface} \> {[-]}\\
{\rm $\delta$} \> {\rm non-dim. film thickness, $h_{0}/D_0$} \> {[-]}\\
{\rm $\delta^{\ast}$} \> {\rm displacement thickness, $0.6479\sqrt{\nu/a}$} \> {[-]}\\
{\rm $\eta$} \> {\rm non-dim. variable, $\sqrt{a/\nu}\;y$} \> {[-]}\\
{\rm $\eta_{max}$} \> {\rm maximum value of $\eta$, $\sqrt{a/\nu}\;h(t)$} \> {[-]}\\
{\rm $\lambda$} \> {\rm impact loss factor, $u_{\infty} /U_0$,} \> {[-]}\\
{\rm $$} \> {\rm $0.26 Re_{0}^{0.05} / (We_{0}^{0.07} \delta^{0.34})$~\cite{Gao2015}} \> {[-]}\\
{\rm $\lambda_1(t)$} \> {\rm time-dependent viscous loss factor} \> {[-]}\\
{\rm $$} \> {\rm $\lambda_1(t) = \bar{u}(t) / u_{\infty}, [$Eq.~(\ref{eq:Hiemz5})]} \> {[-]}\\
{\rm $\lambda_{AG}(t)$} \> {\rm time-dependent combined loss factor,} \> {[-]}\\
{\rm $$} \> {\rm $\lambda_{AG}(t) =\lambda  \lambda_{1}(t)$} \> {[-]}\\
{\rm $\nu$} \> {\rm kinematic viscosity} \> {[$m^2$/$s$]}\\
{\rm $\rho$} \> {\rm density} \> {[$kg$/$m^3$]}\\
{\rm $\sigma$} \> {\rm surface tension} \> {[$Nm^{-1}$]}\\
{\rm $\tau$} \> {\rm non-dim. time, $t U_{0}/D_0$} \> {[-]}\\
{\rm $\tau_0$} \> {\rm non-dim. time of drop impact} \> {[-]}\\
{\rm $\psi$} \> {\rm stream function} \> {[-]}\\
{\rm $\xi$} \> {\rm non-dim. variable, $\eta/\sqrt{t}$ } \> {[-]}\\
\end{tabbing}

\bibliographystyle{abbrv}      
\bibliography{Hiemenz_v3}   

\appendix
\section{Comparison between steady and unsteady stagnation-point flow formulations} \label{App:unsteady}
In this work, the steady, planar stagnation-point flow solution was used for estimating the momentum losses during the propagation of the 
crown. The similarity between the steady planar and axisymmetric stagnation-point flow solution has been already discussed 
in \cite{Schlichting2017} (see pp.~113 - Fig.~5.6) and therefore is not repeated here. The objective of this appendix is to analyse under which conditions unsteady effects play a major role and cannot be neglected in the estimation of momentum losses during the lamella's spreading 
phase. The starting point is represented by the momentum equation in cylindrical coordinates. Neglecting the body force and indicating with 
$u$ and $v$ the velocity component in the radial and $y$ directions, it holds:

\begin{equation}
\frac{\partial u}{\partial t}  +  u \frac{\partial u}{\partial r} + v \frac{\partial u}{\partial y} = \nu \left [  \frac{\partial u}{\partial r} \left( \frac{1}{r} 
\frac{\partial (ur)}{\partial r} \right ) + \frac{\partial^2 u}{\partial y^2}\right ] - \frac{1}{\rho} \frac{\partial p}{\partial r}
\label{eq:aHiemz1}
\end{equation}
The Navier-Stokes equations can be reduced to an ordinary differential equation by the following transformation (axisymmetric SPF) \cite{Drazin2007}:
\begin{equation}
\eta = \sqrt{\frac{a}{\nu}}y,  \quad  \quad f(\eta) = \frac{\psi}{r^2 \sqrt{\nu a}}
\label{eq:aHiemz2}
\end{equation}
where $\psi$ is the stream function. Following Roisman's suggestion \cite{Roisman2009}, self-similar solutions for the unsteady axisymmetric stagnation-point flow can be obtained by eliminating the time dependence with the following additional transformation:
\begin{equation}
\xi = \frac{\eta}{\sqrt{t}} = \sqrt{\frac{a}{\nu t}}y,  \quad  \quad f(\xi) = \frac{g(\xi)}{\sqrt{t}} = \frac{\psi}{r^2 \sqrt{\nu a t}}
\label{eq:aHiemz3}
\end{equation}

Applying the double transformation, the velocity components can be expressed as
\begin{eqnarray*}
u &=& \frac{1}{r} \frac{\partial \psi}{\partial y} = a r f'(\eta) = \frac{a r}{t} g' \left( \sqrt{\frac{a}{\nu t}}y \right ) \\ 
v &=& - \frac{1}{r} \frac{\partial \psi}{\partial r}   = -2 \sqrt{\nu a}  f(\eta) = -2 \sqrt{\frac{a}{\nu t}} g' \left( \sqrt{\frac{a}{\nu t}}y \right ).
\label{eq:aHiemz4}
\end{eqnarray*}

The unsteady term can then be expressed as:
\begin{equation}
\frac{\partial u}{\partial t} = \frac{\partial }{\partial t} \left[ \frac{a r}{t} g' \left( \xi \right ) \right ] = - \frac{a r}{t^2} \left (  \frac{\xi}{2} g'' + g' \right )
\label{eq:aHiemz5}
\end{equation}

The radial pressure gradient can be calculated following the procedure indicated by \cite{Homann1936,Schlichting2017} for an axisymmetric stagnation-point flow of a viscous fluid:
\begin{equation}
\frac{p - p_0}{\rho} = = - \frac{1}{2} a^2 r^2 f'^2 - 2 a \nu (f^2 + f')
\label{eq:aHiemz6}
\end{equation}
Calculating the radial derivative and substituting $f(\xi) = g(\xi)/\sqrt{t}$, it follows:
\begin{equation}
\frac{1}{r} \frac{\partial p}{\partial r}  = = - \frac{a^2 r}{t^2}  g'.
\label{eq:aHiemz7}
\end{equation}
Since the pressure is transmitted integrally through the boundary layer, Eq.~(\ref{eq:aHiemz7}) can be evaluated in the free 
stream, where $g'=1$. Calculating all other derivatives in a similar manner and substituting these variables into Eq.~(\ref{eq:aHiemz1}) yields
\begin{equation}
g''' + 2 g g'' - g'^2 + 1 + \frac{1}{a} \left ( \frac{1}{2}\xi g'' + g' \right ) = 0
\label{eq:aHiemz8}
\end{equation}
with the boundary conditions
\begin{equation}
\xi = 0: \quad g = 0, \, g' = 0; \quad \quad \xi \rightarrow \infty: \quad g' = 1.
\label{eq:aHiemz9}
\end{equation}

\begin{figure}[t]
\centering
{\includegraphics[width=0.75\linewidth]{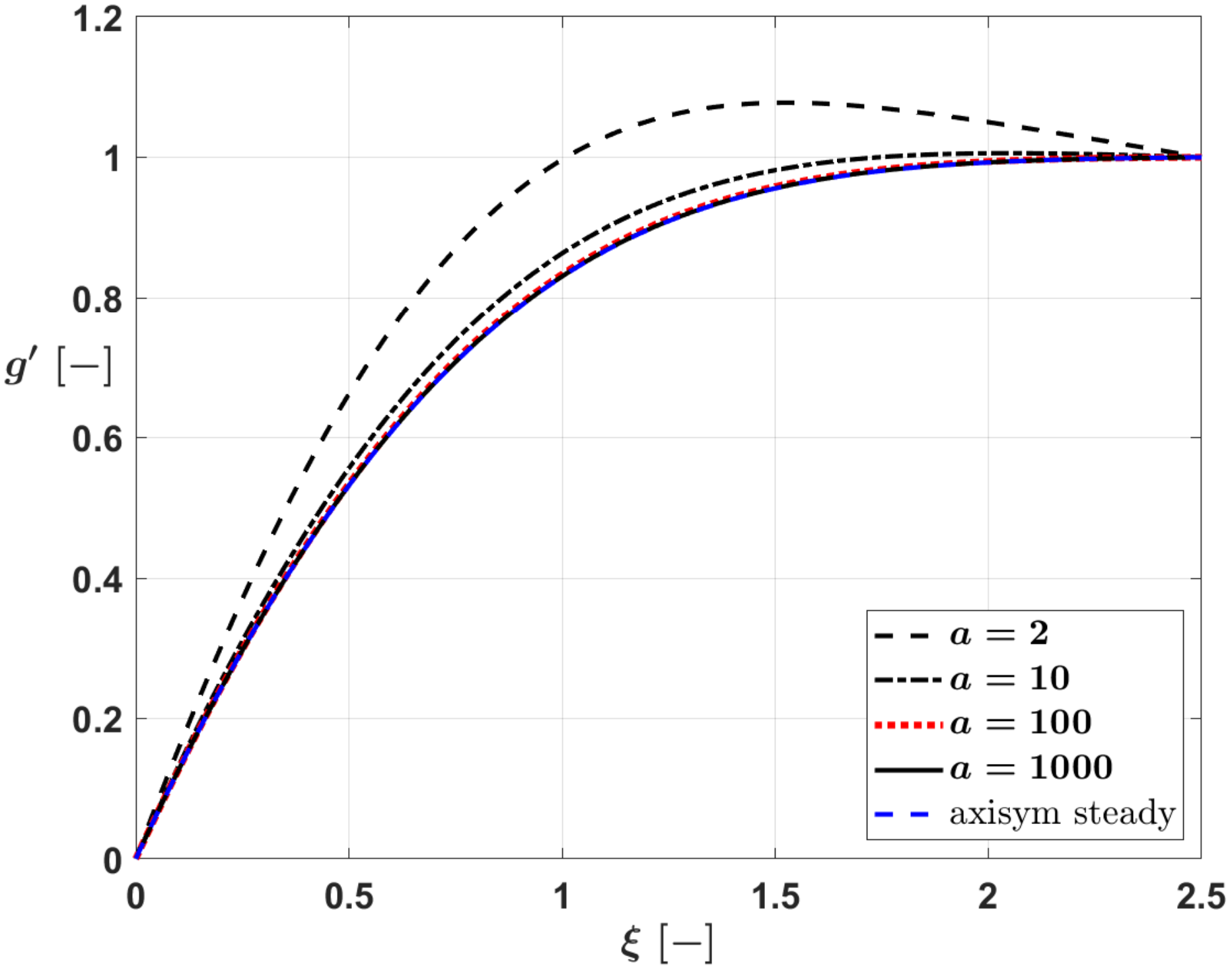}} 
\caption{Self-similar solutions for unsteady axisymmetric stagnation-point flow as function of the parameter $a$, the strength of the potential 
flow. For $a \geq 100$, the unsteady self-similar solution converges towards the steady solution.}
\label{fig:Mod2}
\end{figure}
%
\noindent As can be seen, the unsteady term is represented by  $( 0.5 \, \xi \, g''  +g' )$ and its order of magnitude is weighed by the factor 
$1/ a$. For $a >> 1$, unsteady effects become negligible, in agreement with the analysis presented in \cite{Philippi2016,Riboux2017}, where the 
stagnation-point flow approach was also employed. We solved the 3-ODE for different values of the parameter $a$ and plotted the results 
in Fig.~\ref{fig:Mod2}. Unsteady effects are non-negligible for $a < 10$, e.g.~for a deposition experiment with very low impact velocity against 
a (wetted) wall. For the drop impact cases analysed in the present work, where the parameter $a > 100$ for all experiments, unsteady effects 
are indeed negligible and the solution of Eq.~(\ref{eq:aHiemz8}) converges to the steady solution (blue curve), because the time dependence 
can be completely eliminated from the momentum balance equation.
Finally, as discussed in section \ref{sec:limit}, we recall that both sliding effects and the inception of crown contraction appear to affect the 
accuracy of the model predictions more strongly than unsteady effects.

\section{Dimensionless ejection velocity} \label{App:CBB}
For the evaluation of the dimensionless mass flow rate $Q(\tau)$ [see Eq.~(\ref{eq:res2})], it is necessary to calculated the time derivative of 
the dimensionless ejection velocity $V_s$. Therefore, the dimensionless ejection velocities $V_s(\tau)$ for the \textit{n}-hexadecane case 
($\delta=0.05$, $We_{0}=1384$) and for the hyspin case ($\delta=0.05$, $We_{0}=1423$) are fitted according to the following functional 
dependance:

\begin{equation}
V_{s}(\tau)=b\ exp(c\ \tau)+d\ exp(e\ \tau)
\label{eq:Vs_fit}
\end{equation}

For the corresponding first derivative we obtain therefore:

\begin{equation}
\frac{d}{d\tau} \left[ \frac{1}{V_{s}(\tau)} \right]=-\frac{c\ b\ exp(b\ \tau)+e\ d\ exp(e\ \tau)}{(b\ exp(c\ \tau)+d\ exp(e\ \tau))^2}
\label{eq:dVs_fit}
\vspace{3mm}
\end{equation}

The fitting parameters b,c,d and e are listed in Table~\ref{tab:Fit} for the {n}-hexadecane and hyspin case, respectively.

\begin{table}[h!]
  \centering
  \setlength{\tabcolsep}{3.7pt}  
  \caption{Listing of fitting parameters for $V_s(t)$-fit.}
  \vspace{3mm}
    \begin{tabular}{ccccc}
		 {test liquid} & {b} & {c} & {d} & {e}\\      
      \hline
			{n}-hexadecane & 0.2073 & -0.1826 & 0.3757  & 5.738e-06  \\
			hyspin 				 & 0.3756 & -0.1371 & 0.1096  & -0.01287  \\	
 	\hline
    \end{tabular}
		 \label{tab:Fit}
		\end{table}
%
%
%
%

\section{Details on image processing} \label{app:Exp}
This appendix summarises the details on the imaging technique and the post-processing applied to derive the results presented in the main part of this paper. A comprehensive description can be found in~\cite{Geppert2016,Geppert2017} and~\cite{Geppert2019}.\\

\begin{figure}[t]
\centering
{\includegraphics[width=0.8\linewidth]{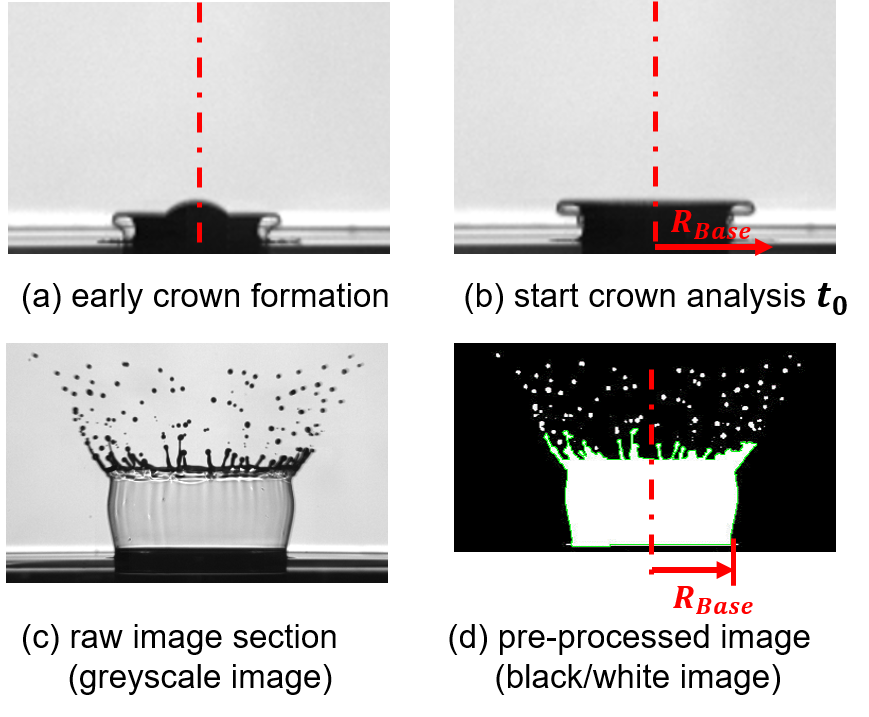}} 
\caption{(a),(b) beginning of crown detection after drop impact; (c),(d) post-processing for crown evaluation.}
\label{fig:CrownDetect}
\end{figure}

\textbf{Image Recording}: A two-perspective high-speed shadowgraphy imaging system is used to simultaneously acquire front and side view 
images of the drop impact. A Photron Fastcam SA1.1 (675K-M1) high-speed camera and a Kern-Paillard YVAR 1:2.8 camera lens with focal 
length of $75$mm are used to record the impact process for $2$s with a frame rate of 20,000 frames per second, a field of view of 896 x 196 
pixels and an effective optical resolution of $80\mu$m/pixel, which corresponds to a magnification factor of 1:4. Shadowgraphy is an appropriate technique for the visualisation of drop impact dynamics~\cite{Castrejon2011} because the large difference in refractive indices between air and 
liquids causes a good contrast on the shadowgrams~\cite{Geppert2019}, as can be seen in Figure~\ref{fig:CrownDetect}c.\\

\textbf{Post-processing}: The post-processing of the recorded images consists of two steps: a pre-processing of the raw images and the evaluation 
of the droplet ($D_0$, $U_0$) and crown properties ($R_{Base}$). The pre-processing converts the raw images, Fig.~\ref{fig:CrownDetect}c, into black/white images, Fig.~\ref{fig:CrownDetect}d, by means of four steps, namely image cropping, normalisation, binarization, and cavity filling. For 
the evaluation of the crown properties, the crown contour (green line, Fig.~\ref{fig:CrownDetect}d), comprising of the outermost white pixels, is determined. The crown's base radius $R_{Base}$ is derived from the corner points of the crown contour's lower part. The complete post-processing 
is automated in Matlab. It analyses both views (front and side) for each image of a drop impact sequence. Thus, the temporal resolution of the analysed data (1/20,000s) corresponds to the image acquisition rate of the camera (20,000 fps).\\

\textbf{Crown detection}:
The beginning of the crown detection is depicted in Figure~\ref{fig:CrownDetect}b, while Figure~\ref{fig:CrownDetect}a shows the situation 
2/20,000s (two frames) earlier. As mentioned in the main part, crown detection starts as soon as the droplet is no longer visible in the images. 
This occurs in between 5 to 10 frames after the first contact between droplet and wall-film, depending on the impact conditions. Only than the automated post-processing routine can detect the crown correctly.

In total, more than 100 drop impact experiments were performed and analysed, covering a wide range of normalised film 
thickness $0.03<\delta<0.55$, droplet Weber number $600<We_{0}<1900$ and droplet Reynolds number $160<Re_{0}<2600$. Please 
note that each experiments was repeated five times to evaluate the reproducibility of the experiments.

\section{Test conditions and fluid properties} \label{app:TC}
\newpage
\begin{table}[t!]
  \centering
  \setlength{\tabcolsep}{3.7pt}  
  \caption{Listing of experimental parameters. More details on the experimental setup can be found in \cite{Geppert2017}. The bold lines indicate that the inception of crown bottom breakup (CBB) was observed in the experiments.}
  \vspace{3mm}
    \begin{tabular}{ccccccccc}
    \multicolumn{9}{l}{Hexadecane:} \\
		\multicolumn{9}{l}{$\sigma = 27.60$ mN/m, $\nu = 4.46$ mm$^2$/s, $\rho=773$ kg/m$^3$} \\
      \hline \hline
      {$D_0$} & {$U_0$} & {$\delta$} & {$h_0$} & ${\lambda}$ & {$a$} & {$We$}   & {$Re$} & {$\delta^{\ast}$}   \\
       {mm} & {$ms^{-1}$} & {-}   & {$\mu m$}  & {-}  & {$s^{-1}$}    & {-}  & {-}                & {$\mu m$}   \\                
      \hline
			2.58 & 3.8 & 0.04 & 110  & 0.65 & 352  & 1043 & 2195 & 73 \\
			2.54 & 4.4 & 0.05 & 130  & 0.62 & 337 & 1384 & 2507 & 75 \\
			2.51 & 4.4 & 0.08 & 200  & 0.54 & 302 & 1349 & 2461 & 79 \\
      2.53 & 4.3 & 0.1  & 253  & 0.51 & 276  & 1314 & 2438 & 82 \\
      2.51 & 4.3 & 0.2  & 500  & 0.40 & 220  & 1304 & 2419 & 92 \\
      2.50 & 4.3 & 0.3  & 750  & 0.34 & 186  & 1299 & 2410 & 100 \\
			2.48 & 4.3 & 0.4  & 1000 & 0.30 & 163  & 1271 & 2374 & 107 \\
      2.50 & 4.3 & 0.5  & 1250 & 0.27 & 146  & 1305 & 2415 & 113 \\
 	\hline
    \end{tabular}
    
    \vspace{3mm}
    \begin{tabular}{ccccccccc}
    \multicolumn{9}{l}{hyspin:} \\
		\multicolumn{9}{l}{$\sigma = 28.65$ mN/m, $\nu = 18.0$ mm$^2$/s, $\rho=878$ kg/m$^3$} \\
      \hline \hline
      {$D_0$} & {$U_0$} & {$\delta$} & {$h_0$} & ${\lambda}$ & {$a$} & {$We$}   & {$Re$} & {$\delta^{\ast}$}   \\
           {mm} & {$ms^{-1}$} & {-}   & {$\mu m$}  & {-}  & {$s^{-1}$}    & {-}  & {-}                & {$\mu m$}   \\                      
      \hline
      \textbf{2.70} & \textbf{4.0} & \textbf{0.03} & \textbf{80}  & \textbf{0.66} & \textbf{312} 	& \textbf{1331} & \textbf{602} & \textbf{156} \\
      \textbf{2.42} & \textbf{4.4} & \textbf{0.05} & \textbf{121} & \textbf{0.58} & \textbf{336}  & \textbf{1423} & \textbf{589} & \textbf{150} \\
      \textbf{2.36} & \textbf{4.4} & \textbf{0.08} & \textbf{180} & \textbf{0.52} & \textbf{301}  & \textbf{1381} & \textbf{573} & \textbf{158} \\
      2.65 & 4.0 & 0.1  & 280 & 0.47 & 227  & 1312 & 592 & 182 \\
      2.57 & 4.3 & 0.2  & 520 & 0.37 & 194  & 1436 & 610 & 197 \\
			2.58 & 4.5 & 0.3  & 800 & 0.30 & 167  & 1601 & 645 & 213 \\
      \hline
    \end{tabular}
    	
	    \vspace{3mm}
    \begin{tabular}{ccccccccc}
    \multicolumn{9}{l}{silicone oil B3:} \\
		\multicolumn{9}{l}{$\sigma = 18.0$ mN/m, $\nu = 3.0$ mm$^2$/s, $\rho=900$ kg/m$^3$} \\
      \hline \hline
      {$D_0$} & {$U_0$} & {$\delta$} & {$h_0$} & ${\lambda}$ & {$a$} & {$We$}   & {$Re$} & {$\delta^{\ast}$}   \\
           {mm} & {$ms^{-1}$} & {-}   & {$\mu m$}  & {-}  & {$s^{-1}$}    & {-}  & {-} & {$\mu m$} \\                      
      \hline
      2.01 & 3.21 & 0.1 & 200  & 0.52 & 262 & 1036 & 2151 & 69    \\
      1.98 & 2.75 & 0.2 & 400  & 0.41 & 182 & 749  & 1815 & 83    \\
      2.00 & 2.61 & 0.3 & 600  & 0.35 & 147 & 681  & 1740 & 93  \\
      \hline
    \end{tabular}

	    \vspace{3mm}
    \begin{tabular}{ccccccccc}
    \multicolumn{9}{l}{silicone oil B10:} \\
		\multicolumn{9}{l}{$\sigma = 20.2$ mN/m, $\nu = 10.0$ mm$^2$/s, $\rho=945$ kg/m$^3$} \\
      \hline \hline
      {$D_0$} & {$U_0$} & {$\delta$} & {$h_0$} & ${\lambda}$ & {$a$} & {$We$}   & {$Re$} & {$\delta^{\ast}$}   \\
           {mm} & {$ms^{-1}$} & {-}   & {$\mu m$}  & {-}  & {$s^{-1}$}    & {-}  & ${-}$                & {$\mu m$}   \\                      
      \hline
      2.01 & 3.58 & 0.1 & 200  & 0.49 & 276 & 1205 & 720 & 123 \\
      1.99 & 3.58 & 0.2 & 400  & 0.38 & 216 & 1193 & 712 & 139 \\
      1.99 & 3.72 & 0.3 & 600  & 0.31 & 187 & 1288 & 740 & 150 \\
      \hline
    \end{tabular}
		  \label{tab:Exp}
			\end{table}
\newpage
\begin{table}[!t]
\centering
    \begin{tabular}{ccccccccc}
    \multicolumn{9}{l}{silicone oil B50:} \\
		\multicolumn{9}{l}{$\sigma = 20.8$ mN/m, $\nu = 50.0$ mm$^2$/s, $\rho=960$ kg/m$^3$} \\
      \hline \hline
      {$D_0$} & {$U_0$} & {$\delta$} & {$h_0$} & ${\lambda}$ & {$a$} & {$We$}   & {$Re$} & {$\delta^{\ast}$}   \\
           {mm} & {$ms^{-1}$} & {-}   & {$\mu m$}  & {-}  & {$s^{-1}$}    & {-}  & ${-}$                & {$\mu m$}   \\                    
      \hline
      2.08 & 4.04 & 0.1 & 210  & 0.44 & 273 & 1567 & 168 & 277 \\
      2.11 & 4.35 & 0.2 & 420  & 0.33 & 219 & 1843 & 184 & 310 \\
      2.09 & 4.35 & 0.3 & 630  & 0.27 & 182 & 1825 & 182 & 340\\
      \hline
    \end{tabular}
\end{table}


\end{document}